\documentclass{article}
\pdfoutput=1
\usepackage[T1]{fontenc}
\usepackage{lmodern}
\usepackage{slantsc}
\usepackage[protrusion=true,expansion=true]{microtype}
\usepackage{breakcites}
\usepackage{multirow}
\usepackage{amsmath,amssymb,amsfonts,amsthm}
\usepackage{subcaption}
\usepackage{graphicx}
\usepackage{fullpage}
\usepackage{setspace}
\usepackage[backref=page]{hyperref}
\usepackage[nameinlink]{cleveref}
\usepackage{color}
\usepackage{wrapfig}
\usepackage{tikz}
\usetikzlibrary{decorations.pathreplacing}
\usepackage{float}
\usepackage{algorithm}
\usepackage[noend]{algpseudocode}

\usepackage[framemethod=tikz]{mdframed}
\usepackage{xspace}
\usepackage{pgfplots}
\usepackage{framed}
\usepackage{thmtools}
\makeatletter
\IfFormatAtLeastTF{2024-11-01}{
  \renewcommand*\@addtoreset[2]{%
    \bgroup
      \edef\aliasctr@@truelist{\aliasctr@follow{#2}}%
      \let\@elt\relax
      \expandafter\@cons\aliasctr@@truelist{{#1}}%
    \egroup
    \expandafter\xdef\csname theH#1\endcsname{%
      \expandafter\noexpand\csname theH#2\endcsname.%
      \noexpand\the\noexpand\value{#1}}%
  }
}{}
\makeatother
\usepackage{thm-restate}
\usepackage{tabu}
\usepackage{fancyhdr}
\pgfplotsset{compat=1.5}

\newtheorem{theorem}{Theorem}[section]
\newtheorem{corollary}[theorem]{Corollary}
\newtheorem{lemma}[theorem]{Lemma}
\newtheorem{proposition}[theorem]{Proposition}
\newtheorem{definition}[theorem]{Definition}

\newtheorem{fact}[theorem]{Fact}

\newenvironment{proofof}[1]{\begin{trivlist} \item {\bf Proof
#1:~~}}
  {\qed\end{trivlist}}

\newcommand{\namedref}[2]{\hyperref[#2]{#1~\ref*{#2}}}

\newcommand{\eps}{\varepsilon}

\newcommand{\mdef}[1]{{\ensuremath{#1}}\xspace}  

\newcommand{\ceil}[1]{\mdef{\left\lceil#1\right\rceil}}               

\definecolor{mahogany}{rgb}{0.75, 0.25, 0.0}
\definecolor{darkblue}{rgb}{0.0, 0.0, 0.55}
\definecolor{darkpastelgreen}{rgb}{0.01, 0.75, 0.24}
\definecolor{darkgreen}{rgb}{0.0, 0.2, 0.13}
\definecolor{darkgoldenrod}{rgb}{0.72, 0.53, 0.04}
\definecolor{darkred}{rgb}{0.55, 0.0, 0.0}
\definecolor{forestgreenweb}{rgb}{0.13, 0.55, 0.13}
\definecolor{greencss}{rgb}{0.0, 0.5, 0.0}
\definecolor{bleudefrance}{rgb}{0.19, 0.55, 0.91}

\newcommand{\ignore}[1]{}

\newif\ifnotes\notestrue 
\ifnotes
\newcommand{\samson}[1]{\textcolor{blue}{{\bf (Samson:} {#1}{\bf ) }} \marginpar{\tiny\bf
             \begin{minipage}[t]{0.5in}
               \raggedright S:
            \end{minipage}}}
\newcommand{\samsonz}[1]{\textcolor{red}{{\bf (Samson:} {#1}{\bf ) }} \marginpar{\tiny\bf
             \begin{minipage}[t]{0.5in}
               \raggedright S:
            \end{minipage}}}
\newcommand{\neeraj}[1]{\textcolor{purple}{{\bf (Neeraj:} {#1}{\bf ) }} \marginpar{\tiny\bf
             \begin{minipage}[t]{0.5in}
               \raggedright N:
            \end{minipage}}} 
\newcommand{\katherine}[1]{\textcolor{forestgreenweb}{{\bf (Katherine:} {#1}{\bf ) }} \marginpar{\tiny\bf
             \begin{minipage}[t]{0.5in}
               \raggedright K:
            \end{minipage}}} 
\else
\newcommand{\samson}[1]{}
\newcommand{\samsonz}[1]{}
\newcommand{\neeraj}[1]{}
\newcommand{\katherine}[1]{}
\fi

\makeatletter
\renewcommand*{\@fnsymbol}[1]{\textcolor{mahogany}{\ensuremath{\ifcase#1\or *\or \dagger\or \ddagger\or
 \mathsection\or \triangledown\or \mathparagraph\or \|\or **\or \dagger\dagger
   \or \ddagger\ddagger \else\@ctrerr\fi}}}
\makeatother

\providecommand{\email}[1]{\href{mailto:#1}{\nolinkurl{#1}\xspace}}

\hypersetup{
     colorlinks   = true,
     citecolor    = mahogany,
	 linkcolor	  = darkpastelgreen
}

\fancypagestyle{pg}
{
\lhead{}
\rhead{}
\cfoot{--\ \thepage\ --}

}

\AtBeginDocument{%
  \DeclareFontShape{T1}{lmr}{m}{scit}{<->ssub*lmr/m/scsl}{}%
}
\usepackage{graphicx}
\usepackage{tabularx}
\title{A Simple Black-Box Framework for Differentially Private Continual Release on Data Streams}

\author{
Neeraj Gogate\thanks{Texas A\&M University.
E-mail: \email{neerajgogate@tamu.edu}.}
\and
Katherine Tomashevsky\thanks{Texas A\&M University.
E-mail: \email{ktomashe@tamu.edu}.}
\and
Samson Zhou\thanks{Texas A\&M University.
E-mail: \email{samsonzhou@gmail.com}.
Samson Zhou is supported in part by NSF CCF-2335411. 
}
}

\usepackage{booktabs}
\usepackage{longtable}
\usepackage{array}

\begin{document}
\allowdisplaybreaks
\maketitle
\begin{abstract}
We give a simple black-box transformation for differentially private continual release of streaming statistics. Our framework applies to any nonnegative, monotone statistic on insertion-only streams with bounded sensitivity and polynomially bounded range, provided there is a tunable streaming algorithm that approximates the statistic at every time. The transformation runs a single instance of the underlying streaming algorithm and privately detects crossings of geometrically spaced thresholds using sparse vector. A scale-dependent clipping argument controls the sensitivity of the approximate estimates without requiring access to the internal structure of the streaming algorithm. For a stream of length $m$ and target multiplicative error $\gamma$, the resulting mechanism is $(\varepsilon,\delta)$-differentially private and uses only $O(\log m/\gamma)$ additional space beyond the base algorithm. For polynomially bounded statistics, it achieves additive error
$\widetilde{O}((\Delta_F+\kappa)\log m/(\varepsilon\gamma)+\kappa)$
while requiring the base algorithm to run at a multiplicative accuracy smaller than $\gamma$ by only polylogarithmic factors. Thus, for many standard streaming algorithms, the transformation preserves sublinear space while providing privacy for the entire sequence of outputs. We instantiate the framework for distinct elements, frequency norms, longest increasing subsequence, regression and low-rank approximation loss, clustering cost, and union volume estimation.
\end{abstract}
\newpage

\section{Introduction}
Streaming algorithms provide a powerful framework for processing massive datasets using memory far smaller than the input itself~\cite{AlonMS99,muthukrishnan2005data}. 
Given a stream of updates $e_1,\ldots,e_m$, the goal at each time $i\in[m]$ is to estimate a statistic $F_i(D)=F(e_1,\ldots,e_i)$ of the current dataset, while using space sublinear in the stream length or universe size. 
Since many fundamental statistics require linear space for exact computation, streaming algorithms instead maintain an approximation $\widehat{F}_i$ obeying a multiplicative-additive guarantee
$$(1-\gamma) F_i(D)-\kappa \leq \widehat{F}_i(D) \leq (1+\gamma)F_i(D)+\kappa,$$
where $\gamma$ is the multiplicative error and $\kappa$ is the additive error. An algorithm which obeys this guarantee simultaneously for all times $t\in[m]$ with probability at least $1-\delta$ is called \textit{($\gamma,\kappa,\delta$)-continually accurate}.

Although a standard streaming algorithm may only be required to return an estimate after the full data stream, many applications require constant monitoring, tracking the desired statistic over the course of the entire stream. 
This is the continual release setting, in which an algorithm outputs an estimate after each update and its accuracy must hold at all $m$ times. 
A \emph{strong tracker} satisfies this condition through a single high-probability event that guarantees accuracy at every time simultaneously. 
There are known strong trackers for many fundamental problems, including distinct elements and frequency norms~\cite{BravermanCIW16,blasiok_et_al:LIPIcs.APPROX-RANDOM.2017.32,10.1145/3309193}.

\paragraph{Differential privacy and continual release.}
Many applications of continual release involve data streams containing sensitive information, such as healthcare records, financial transactions, or user activity. 
In these settings, the sequence of released estimates itself can reveal private information, and protecting only individual outputs is insufficient. 
We thus focus on the continual-observation model of \cite{dwork2010differential,chan2011private}, adopting differential privacy as the privacy requirement. 
We call two streams \emph{neighboring} if they differ in a single update, corresponding to the contribution of one individual. 
For a randomized continual-release mechanism $\mathcal{M}$ whose output is the transcript $\mathcal{M}(D)=(\bar{Y}_1,\dots,\bar{Y}_m)$, we say that $\mathcal{M}$ is $(\varepsilon,\delta)$-differentially private~\cite{DworkMNS06} if, for every pair of neighboring streams $D\sim D'$ and every set $\mathcal{S}$ of possible output transcripts,
$$\Pr[\mathcal{M}(D)\in\mathcal{S}]
\leq e^{\varepsilon}\Pr[\mathcal{M}(D')\in\mathcal{S}]+\delta.$$ 
Thus, differential privacy constrains the joint distribution for all $m$ releases, rather than protecting each output in isolation. 
Achieving privacy for the full transcript is substantially more challenging than ensuring a single output is private, since the mechanism releases information at every time step and na\"{i}ve composition can cause privacy loss to grow polynomially with the stream length $m$.

Ideally, adding privacy guarantees to a streaming algorithm would require no knowledge of its internal structure: one would simply transform any non-private streaming algorithm into a private one while preserving its accuracy guarantees. 
A natural first attempt would be to add noise calibrated to the sensitivity $\Delta_F$ of the true statistic. 
However, this approach breaks down for approximation algorithms, since the sensitivity of $F$ does not imply a comparable sensitivity bound for the algorithmic estimate $\widehat{F}$. 
For neighboring streams $D\sim D'$ satisfying $|F_i(D)-F_i(D')|\leq \Delta_F$, the two estimates may lie at opposite ends of their respective approximation intervals, so that
$$|\widehat{F}_i(D)-\widehat{F}_i(D')|\leq 2\gamma F_i(D)+(1 + \gamma)\Delta_F+2\kappa,$$
which can grow with the magnitude of the statistic. 
Thus, calibrating noise to the sensitivity of $F$ may fail to provide privacy, while calibrating to the worst-case sensitivity of $\widehat{F}$ can destroy utility. 
This tension raises the central question of our work:

\begin{quote}
{\textit{Under what conditions can a non-private continual streaming approximation algorithm be transformed, in a black-box manner, into a differentially private continual-release algorithm while preserving meaningful accuracy and sublinear space?}}
\end{quote}

\subsection{Main Result}
Our main contribution is a unified black-box transformation. Any tunable and continually accurate streaming algorithm that approximates a nonnegative, monotone, insertion-only function with bounded range can be transformed into a differentially private continual release algorithm while retaining tunable multiplicative accuracy and adding only $O((\log M+1)/\gamma)$ space. Rather than designing a separate private mechanism for each streaming problem, the same threshold based wrapper applies whenever these structural conditions hold. The construction uses only one instance of the underlying streaming algorithm, geometrically spaced thresholds, clipping, and standard \textsc{AboveThreshold} instances. The corresponding analysis reduces to controlling the sensitivity at each threshold scale and sandwiching the released threshold around the true value.

The formal notions used in the theorem, including continual accuracy and tunability, are defined in \Cref{sec:preliminaries}. We state the main consequence here directly in terms of the desired privacy, accuracy, and space guarantees; the exact parameter choices used by the construction are deferred to \Cref{thm:main}.

\begin{theorem}
\label{thm:main:simp}
Let $F$ be a nonnegative, monotone, insertion-only streaming statistic with sensitivity $\Delta_F$ and range bounded by $M\geq 1$. Suppose there is a tunable monitoring algorithm $\mathcal A$ that, for parameters $(\alpha,\kappa,\delta)$, is $(\alpha,\kappa,\delta)$-continually accurate and uses $S_{\mathcal A}(m,\alpha,\kappa,\delta)$ space.

Then, for any $\gamma,\varepsilon\in(0,1]$, $\delta\in(0,1)$, and $\kappa\geq 0$, there is a privatized algorithm $\mathcal A'$ satisfying:
\begin{enumerate}
\item \textbf{Privacy.} $\mathcal A'$ is $(\varepsilon,\delta)$-differentially private.
\item \textbf{Accuracy.} $\mathcal A'$ is $(\gamma,\rho,\delta)$-continually accurate, where

$$\rho
:=
\frac{36(\Delta_F+2\kappa)}{\varepsilon\gamma}
(\log M+1)
\log\left(
    \frac{6(m+1)(\log M+1)}{\gamma\delta}
\right)
+2\kappa+2.$$

\item \textbf{Space.} $\mathcal A'$ uses space at most

$$S_{\mathcal A}\left(
    m,
    \frac{\varepsilon\gamma^2}
    {288(\log M+1)
    \log\left(
        \frac{6(m+1)(\log M+1)}{\gamma\delta}
    \right)},
    \kappa,
    \frac{\delta}{2}
\right)
+
O\left(\frac{\log M+1}{\gamma}\right).$$

\end{enumerate}
\end{theorem}
A useful feature of the theorem is that the proof is largely independent of the particular statistic being tracked. Once a suitable non-private continual approximation algorithm is available, privatization does not require a new problem-specific mechanism or privacy proof: the same wrapper and analysis can be reused. We therefore view the breadth of the result as part of its value, and illustrate it through a collection of applications below.

\subsection{Applications}
Our framework yields private continual-release guarantees for a broad range
of streaming problems by applying the same black-box transformation to
existing continual approximation algorithms.
\Cref{sec:applications-main} presents representative guarantees
for distinct elements, $L_p$ norm estimation with $p\in[1,2]$,
and longest increasing subsequence, and discusses their
relationship to existing private algorithms.
Full tables comparing privacy, accuracy, and space guarantees
with non-private base algorithms and application-specific
private methods appear in Appendix~\ref{app:comparisons}.
Complete statements and proofs appear in
the rest of Appendix~\ref{app:applications}, including additional ranges
of $p$, regression and low-rank approximation loss,
clustering cost, and union volume estimation.

\subsection{Framework Overview}

Our framework takes a continuous monitoring streaming algorithm $\mathcal{A}$ for a statistic $F$ and wraps it in a private threshold-crossing mechanism. The same construction can therefore be applied to a broad set of existing algorithms, rather than be rederived for each statistic.
The central idea is simple: we only need to release a new estimate of the statistic when our previous estimate is at risk of being inaccurate. 
We place a geometric ladder of thresholds
$$T_j = (1 + \gamma/2)^j, \qquad j = 0,1, \cdots, N-1, \qquad N = O\left(\frac{1}{\gamma} \log{M}\right)$$
across the range $[1, M]$. 
For each threshold, we run an independent instance of \textsc{AboveThreshold} (sparse-vector), which privately detects when the estimate first reaches that scale, and at every time we output the largest threshold detected so far. 
Here we require a monotonic statistic, so that each level only needs to be detected once, and the geometric spacing provides a $(1+\gamma)$-approximation to the current value. 
The base algorithm is used purely as a black box, accessed only through its current numerical estimate and never inspected or modified. 
The entire mechanism consists of a single pass of the algorithm together with $N$ constant size-threshold detectors, adding $O(N)$ space and $O(N)$ work per update.

\paragraph{Approximation sensitivity.}The main complication is that the output of an approximation algorithm need not inherit the low sensitivity of the statistic it approximates. 
Even when neighboring streams satisfy $| F_i(D) - F_i(D')| \leq\Delta_F$, two valid approximations can lie near opposite ends of their ``allowable'' intervals, e.g., $\widehat{F}_i(D)\approx(1-\gamma){F}_i(D)$, while $\widehat{F}_i(D')\approx(1+\gamma){F}_i(D')$, so that the sensitivity of the output can be as large as $O(\gamma F_i(D))$. 
Thus, although the true statistic may have small sensitivity, feeding the raw estimate $\widehat{F}_i$ to \textsc{AboveThreshold} directly can require noise that grows with the value of the statistic.

We address this issue with scale-dependent clipping. 
At level $j$, we test the clipped query $\min\{\widehat{F}_i, R_j\}$ rather than $\widehat{F}_i$ itself, where $R_j$ can be viewed as a radius that is large enough that clipping does not affect the above-threshold certificate used for accuracy, yet small enough to admit useful privacy noise. 
Consequently, conditioned on the event that the base streaming algorithm is simultaneously accurate on both neighboring streams, each threshold instance receives a low-sensitivity query at its own local scale rather than paying for the worst-case sensitivity of the approximation over the entire range. 

To illustrate this, consider estimating the number of distinct elements in a stream using an approximation algorithm with a multiplicative error bound of $\gamma = \frac{1}{4}$. Suppose the stream elements $e_i$ are drawn from the universe $[100]$. The global sensitivity of the estimator $\widehat{F}$ can exceed $50$, even though the true distinct elements count $F$ has a sensitivity of $\Delta_F = 1$. If we query the SVT instances using geometrically spaced thresholds $T_j = 2^j$, calibrating the noise to this worst-case sensitivity of $50$ precludes any multiplicative accuracy guarantees for smaller thresholds (e.g., $T_j \in \{1, 2, 4, \dots\}$). However, by utilizing modified thresholds $R_j = \frac{4}{3}T_j$, the noise scales with $\frac{2}{3}T_j$. 
This ensures the noise remains proportional to the specific threshold being tested rather than being dominated by the maximum global sensitivity.

\paragraph{Accuracy.}The accuracy argument is then a sandwiching argument. 
If the mechanism currently releases threshold $T_j$, the $j$-th level has been crossed while the next level has not. 
The accuracy guarantee of \textsc{AboveThreshold} places the clipped estimate between the corresponding upper and lower margins, and the guarantee of the base algorithm converts these certificates back into bounds on the true statistic. 
Because consecutive thresholds are geometrically separated, this yields the desired multiplicative approximation; the sparse-vector noise and the base algorithm additive error contribute only to the final additive term. 

\paragraph{Pure privacy for randomized algorithms.}
For randomized base algorithms, the clipped-query sensitivity bound holds
on a high-probability accuracy event and accounting for its failure yields
the additive $\delta$ in our privacy guarantee. For deterministic base
algorithms satisfying the accuracy guarantee on every input, the wrapper
already provides pure $\varepsilon$-differential privacy.
Appendix~\ref{sec:pure-dp} gives a pure-private variant for randomized
algorithms using a finite-output-range conversion, together with its
accuracy and space guarantees.

\subsection{Related Work}

\paragraph{Differential privacy under continual observation.}
The continual-observation model was introduced independently by Dwork et al.~\cite{dwork2010differential} and Chan et al.~\cite{chan2011private}. 
Their hierarchical mechanisms privately release running counts by decomposing the time horizon into intervals and adding noise to the associated partial sums. Subsequent work has refined the accuracy and efficiency of continual counting, including matrix-mechanism and factorization-based approaches that obtain nearly optimal error for prefix sums and extend to more weighted or decaying sums \cite{henzinger2023almost,henzinger2024unifying}. These techniques fundamentally exploit the linear structure of the released queries. In contrast, the statistics considered here may be nonlinear and may be available only through an approximation algorithm.
The need for additional structure beyond low sensitivity is reflected in the lower bounds of Jain et al.~\cite{jain2023price}, who exhibit natural problems for which continual release requires polynomially more error than the corresponding batch problem. Separate works address turnstile streaming problems, such as distinct elements and moment estimation, achieving $\operatorname{polylog}$ error and space requirements in a continual release setting \cite{jain2023counting,aamand2026skirting}. 
We address these problems in our applications; however, our approach differs in that we can use a black-box transformation at the expense of requiring an insertion only setting. Our assumptions of monotonicity, polynomially bounded range, and access to a tunable continual approximation restrict the framework to a structured class of statistics for which threshold crossings can be used to control the information released over time. 

\paragraph{Geometric thresholding for monotone statistics.}
Using sparse vector with geometrically increasing thresholds to obtain
tunable multiplicative error is an established technique;
Fichtenberger, Henzinger, and Ost~\cite{fichtenberger2021continual}
give a general continual-observation mechanism for monotone functions
with bounded sensitivity, using this approach to obtain
multiplicative-additive accuracy.
Our contribution extends this approach to statistics with
a tunable streaming approximation algorithm.
The approximation therefore need not inherit the sensitivity of the
underlying statistic, so we use scale-dependent clipping and an analysis
over good random seeds to control their differences on neighboring streams.

\paragraph{Problem-specific private streaming algorithms.} A separate line of work develops private continual-release algorithms for particular streaming problems rather than for prefix sums. Epasto et al.~\cite{epasto2023differentially} give continual-release algorithms for frequency moments, together with reductions for primitives including distinct elements, heavy hitters and low-frequency elements; they also extend their techniques to sliding windows. 

The literature also considers update models that are more general than the monotone insertion-only setting we study here. 
Jain et al.~\cite{jain2023counting} study continual release of distinct elements in turnstile streams, where insertions and deletions are both permitted, and show a polynomial additive-error barrier together with parameterized guarantees in terms of the stream's flippancy. Aamand et al.~\cite{aamand2026skirting} subsequently show that allowing simultaneous multiplicative and additive error can circumvent this barrier for distinct elements and $F_2$ while retaining polylogarithmic space. 
Differential privacy in sliding-window and other private streaming models has likewise been studied for heavy hitters~\cite{upadhyay2019sublinear}, graph streams~\cite{upadhyay2021differentially}, and matrix problems such as spectral approximation, PCA, and regression~\cite{upadhyay2021framework}. 
These results are complementary to ours. 
In turnstile and sliding-window streams, the target statistic need not be monotone over time, so the sandwiching argument used by our transformation does not directly apply. 
Conversely, by restricting attention to monotone insertion-only statistics we can avoid designing a new private mechanism for each individual problem and instead inherit guarantees from a non-private continual approximation algorithm.

\paragraph{Black-box privatization of approximation algorithms.}
Generic transformations have also been studied for data that arrive over time. Cummings et al.~\cite{cummings2018differential} show that private algorithms for static databases can be scheduled and rerun at selected points to obtain private algorithms for growing databases with little loss in accuracy. Their transformation starts from an already-private static algorithm and addresses adaptive analysis of a growing database, rather than privatizing the complete prefix-by-prefix transcript of a non-private streaming approximation algorithm.

The closest conceptual predecessor to our work is the black-box transformation of Blocki et al.~\cite{blocki2023makeapproximationalgorithmprivate}. Their framework transforms a tunable multiplicative-additive approximation algorithm for a low-sensitivity function into a differentially private approximation algorithm using smooth sensitivity. It applies to sublinear-time and streaming algorithms for problems including $L_p$-norm estimation, distinct elements, minimum spanning tree weight, and sliding-window statistics.
However, the guarantee of Blocki et al. is one-shot, including in their streaming applications. The transformed algorithm privately releases the value of the statistic on the dataset represented by the stream, rather than protecting output after every prefix. 
Consequently, their theorem cannot simply be invoked on a strong tracker to obtain continual observation. 
Running their one-shot private mechanism separately at every time step would require composing privacy over $m$ prefix datasets and dividing the privacy budget among many releases, which substantially degrades the resulting accuracy. 
Treating the entire output transcript as one vector-valued query also loses the scalar low-sensitivity structure used in their smooth-sensitivity analysis: one changed update can affect every subsequent prefix and hence many coordinates of the transcript.

Our framework addresses this composition problem. We run a single non-private continual approximation algorithm but reveal only which geometrically spaced levels have been reached. Sparse-vector instances privately detect these threshold crossings, and clipping localizes the sensitivity calculation to the scale of the active threshold.  As a result, the mechanism spends privacy on a logarithmic number of monotone progress events rather than on all $m$ estimates.  Thus Blocki et al. and the present work are complementary black-box transformations: the former privatizes a single approximate answer, while ours privatizes an entire monotone streaming trajectory under additional structural assumptions that make continual release possible.

\section{Preliminaries}
\label{sec:preliminaries}
Fix a universe $\mathcal{U}$ and a horizon $m$. A stream is $D := (e_1,\cdots, e_m) \in (\mathcal{U} \cup \{\bot\})^m$, where the null update has no effect. We write $D_{\leq i} := (e_1,\cdots, e_i)$ for the first $i$ updates and $F_i(D) = F(D_{\leq i})$. Two streams $D, D'$ are \emph{neighboring} if they differ in one coordinate and one of the two entries at that coordinate is $\bot$. This formulation represents insertion or deletion of one event without shifting later time indices.

\begin{definition}[Monotone Statistic]
A streaming statistic $F$ is \emph{monotone} on insertion-only streams if $F_1(D) \leq F_2(D) \leq ... \leq F_m(D)$ for every insertion-only stream $D$.
\end{definition}
The sensitivity of a function bounds how much changing one element of a dataset can affect the value of the function. This quantity is heavily relied upon in differential privacy, as privatization aims to produce outputs whose distributions are nearly identical for neighboring datasets. Thus, it is standard for noise added in privatization to be proportionate to sensitivity.
\begin{definition}[Sensitivity]
$F$ has sensitivity $\Delta_F$ if, for every pair of neighboring streams, $D \sim D'$ and every time $i$, $|F_i(D) - F_i(D')| \leq \Delta_F.$
\end{definition}
Throughout our analysis, we assume that the range of $F$ is bounded by a known finite upper-bound $M$. However, in many applications $M$ is polynomial in $m$, and allows us to further bound results with $\log M=O(\log m)$.
\begin{definition}[Polynomial-bounded range]
The range of statistic $F$ is polynomial-bounded if there exists constant $c$ such that $F_i(D)\leq m^c$ for every stream $D$ and all times $i$. This assumption is used throughout our applications to tighten space and accuracy bounds.
\end{definition}
\begin{definition}[Tunable continual approximation]
A streaming algorithm $\mathcal{A}$ is a tunable continual approximation algorithm for $F$ if, for every choice of parameters $\alpha \in (0,1],\kappa \geq 0,\delta \in (0,1)$, it outputs values $\widehat{F}_i(D)$ satisfying
$(1-\alpha)F_i(D)-\kappa \leq \widehat{F}_i(D) \leq (1+\alpha)F_i(D)+\kappa$
 for every time $i \in [m]$
with probability at least $1 - \delta$.
\end{definition}
\begin{definition}[Continual accuracy]
A randomized release sequence
$\bar{Y}_1,\dots,\bar{Y}_m$ is $(\gamma, \rho, \delta)$-continually accurate on $D$ if, with probability at least $1-\delta$, simultaneously for every time $i$,
$
(1-\gamma)F_i(D)-\rho
\leq
\bar{Y}_i
\leq
(1+\gamma)F_i(D)+\rho,
$
where $F_i$ is the true value of the function being approximated at time $i$.
\end{definition}
\begin{definition}[Differential Privacy]
A randomized algorithm $\mathcal{M}$ is $(\varepsilon,\delta)$-differentially private (DP) if for all neighboring datasets $D\sim D'$, and all measurable subsets $S \subseteq \text{Range}(\mathcal{M})$,
$\Pr[\mathcal{M}(D) \in S] \leq e^{\varepsilon} \Pr[\mathcal{M}(D') \in S] + \delta$

\end{definition}

\begin{definition}[Continual Release Differential Privacy]
Let $\mathcal M(D)=(Y_1,\ldots,Y_m)$ denote the complete transcript
released on stream $D$. The mechanism $\mathcal M$ is
$(\varepsilon,\delta)$-differentially private if, for every pair of
neighboring streams $D\sim D'$ and every measurable set $S$ of output
transcripts,
$\Pr[\mathcal{M}(D) \in S] \leq e^{\varepsilon} \Pr[\mathcal{M}(D') \in S] + \delta.$
\end{definition}
\begin{definition} [Soft-O notation]
    Let $f:D\to\mathbb{R}$ and $g:D\to\mathbb{R}$. Then $f(x)=\widetilde{O}(g(x))\iff f(x)=O\left(g(x)\operatorname{polylog}(g(x))\right)$
\end{definition}
\paragraph{Space model.}
Unless a bound is explicitly stated in bits, space is measured in machine words. The $O(N)$ overhead of the wrapper counts $O(1)$ words for each threshold instance under the real-RAM convention used for the Laplace samples in \textsc{AboveThreshold}. When a cited base algorithm is stated in bits, we preserve that unit in the corresponding application statement.

We may assume without loss of generality that the stated base-algorithm
space bound $S_{\mathcal A}(m,\alpha,\kappa,\delta)$ is non-increasing
in $\alpha$. If a stated bound is not presented in this form, we replace
it by its monotone envelope. Thus, whenever the wrapper runs the base
algorithm with $\alpha'\geq\alpha$, its space is at most
$S_{\mathcal A}(m,\alpha,\kappa,\delta)$.

\paragraph{AboveThreshold.}
We use the standard one-answer sparse-vector primitive
\textsc{AboveThreshold}~\cite{dwork2014algorithmic}. Given a threshold $T$, a privacy parameter $\varepsilon^{\prime}$, and deterministic queries $q_1, \cdots, q_m$ of sensitivity at most $\Delta$, the mechanism samples $\widehat{T} = T + \operatorname{Lap}(2\Delta/\varepsilon^{\prime})$ and at time $i$, compares $q_i + \operatorname{Lap}(4 \Delta/\varepsilon^{\prime})$ with $\widehat{T}$. It outputs $\bot$ until the first crossing, outputs $\top$ at that crossing and then halts.

We use two familiar properties of this primitive. First, its complete output is $\varepsilon^\prime$-differentially private. Second, for any failure probability $\delta^\prime \in (0,1)$, with probability at least $1 - \delta^\prime$, simultaneously over all processed queries,
\begin{enumerate}
    \item If the mechanism has not fired by time $i$, then $q_i < T + \alpha_{\mathrm{AT}}.$
    \item If the mechanism fires at time $k$, then $q_k \geq T - \alpha_{\mathrm{AT}}.$
\end{enumerate}
where $\alpha_{\mathrm{AT}} = \frac{6\Delta}{\varepsilon^{\prime}} \log{\left( \frac{m+1}{\delta^\prime}\right)}$
We give formal statements and proofs for these properties of \textsc{AboveThreshold} in \Cref{app:above-thresh}.

\paragraph{Commonly used parameters} The following serves as a reference for parameters and variables used throughout our analysis:  
\begin{center}
\renewcommand{\arraystretch}{1.18}
\begin{tabularx}{\linewidth}{@{}l l >{\raggedright\arraybackslash}X@{}}
\toprule
Parameter & Value Bound & Description \\
\midrule
$\gamma$ & $(0,1]$ & Desired multiplicative error of the final release \\
$\kappa$ & $\mathbb R^{\geq 0}$ & Additive error supplied to the base algorithm \\
$\delta$ & $(0,1)$ & Target for privacy slack and total accuracy failure \\
$\varepsilon$ & $\mathbb R^{+}$ & Total privacy budget \\
$m$ & $\mathbb N$ & Stream length \\
$M$ & $\operatorname{poly}(m),\mathbb R^{+}$ & Upper bound on values of $F$ \\
$\mathcal A$ & Base algorithm & Underlying streaming algorithm \\
\bottomrule
\end{tabularx}
\end{center}
For notational simplicity, the main theorem uses the same top-level parameter $\delta$ for the approximate-DP slack and the overall accuracy-failure probability. Internally, we run the base algorithm with failure probability $\delta_A:=\delta/2$ and assign each threshold test failure probability $\delta_{\mathrm{AT}}:=\delta/(2N)$.

\section{Framework and Analysis}
In this section, we give the formal specification of our framework \textsc{PrivateStream}, which uses a black box streaming algorithm to produce function approximations and releases privatized approximations at each time. 
After each update, we feed the approximation into a series of SVT instances with geometrically increasing thresholds. 
When a new threshold is surpassed, we release this threshold as the private approximation. We maintain accuracy by using threshold spacing dependent on desired multiplicative accuracy $\gamma$ and limiting sensitivity relative to threshold value. 

We first fix the threshold and budget parameters, then define the clipping radii and state the mechanism.  The sensitivity, privacy, and accuracy arguments are deferred to \Cref{sec:privacy,sec:accuracy}.

\paragraph{Thresholds and budget application.}
Fix privacy parameters $\varepsilon > 0$ and $\delta \in (0,1)$ and target multiplicative error $\gamma \in (0,1].$ Fix the threshold spacing parameter $b$, each threshold value $T_j$, the index of the last threshold $L$, and the total number of SVT instances $N$ as follows: 
$$b = \frac{\gamma}{2}, \qquad T_j := (1+b)^j,\qquad L := \lceil \log_{1+b}{M} \rceil, \qquad N = L + 1.$$ 
Thus the thresholds $T_0 , \cdots , T_L$ cover $[1,M]$ using $N = O(\gamma^{-1}\log M + 1)$ levels. We split the privacy budget uniformly across the $N$ instances and split the accuracy-failure budget between the base algorithm and the threshold tests, where each SVT instance recieves privacy budget $\varepsilon'$ and accuracy failure budget $\delta'$:

$$\varepsilon^{\prime} := \frac{\varepsilon}{N}, \qquad \delta' = \frac{\delta}{2N} \qquad H := \log\!\left(\frac{m+1}{\delta^{\prime}}\right) = \log\!\left(\frac{2(m+1)N}{\delta}\right).$$
Here $H$ is the Laplace tail parameter needed for simultaneous accuracy over all times and thresholds. 

\paragraph{Tracker accuracy and clipped queries.}
We run the base algorithm $\mathcal{A}$ with internal multiplicative accuracy
$\tau := \min\left\{\frac{\gamma}{16}, \frac{\varepsilon \gamma}{96NH}\right\}.$
For a fixed sequence of random coins $r$, let
$$\mathcal{E}_{\mathcal{A}}(D:r) := \left\{(1-\tau)F_i(D) - \kappa \leq \widehat{F}_i(D) \leq (1+\tau)F_i(D) + \kappa \text{ for every } i \right\}.$$
By the base algorithm guarantee, $\Pr[\mathcal{E}_{\mathcal{A}}(D;r)] \geq 1 - \delta/2$ for every stream $D$. 
At level $j$, the mechanism gives the clipped queries $q_{i,j}(D;r) := \min\{ \widehat{F}_i(D;r), R_j\}$ to \textsc{AboveThreshold}. 
The sensitivity calculation in \Cref{lem:sensitivity} shows that whenever the base algorithm is simultaneously accurate on both neighboring streams, these queries have sensitivity at most
$$\Delta_j := \frac{2\tau}{1 + \tau}R_j+ (1 - \tau)\Delta_F + \frac{2\kappa}{1 + \tau}.$$
With failure probability $\delta'$, the standard accuracy guarantee for \textsc{AboveThreshold} therefore incurs additive error
$\alpha_j := \frac{6H \Delta_j}{\varepsilon^{\prime}}.$
\paragraph{Clipping radii.}
We choose $R_j$ to be the smallest radius for which clipping does not interfere with this guarantee, namely the least solution to $R_j  =T_j + \alpha_j.$
To write this solution explicitly define
$\lambda := \frac{12\tau NH}{(1 + \tau)\varepsilon}$ and $\eta := \frac{6NH}{\varepsilon}\left((1 - \tau)\Delta_F + \frac{2\kappa}{1 + \tau}\right).$
These parameters can be used to rearrange $\alpha_j$ to be explicitly dependent on $T_j$ and establish a minimal clipping radius. 
$\alpha_j = \lambda R_j + \eta$ and hence
$R_j = \frac{T_j + \eta}{1 - \lambda}.$
The choice of $\tau$ implies $\lambda \leq \gamma/8 < 1$, so these quantities are well defined.
\begin{proposition} [Fixed point clipping]
\label{prop:optimal-clipping}
For every $j$, $\alpha_j = \lambda R_j + \eta = R_j - T_j = \frac{\lambda T_j + \eta}{1- \lambda}.$ In particular, $R_j = T_j + \alpha_j$ is the smallest clipping radius that covers the level-$j$ \textsc{AboveThreshold} error.
\end{proposition}
\begin{proof}
Substituting the definition of $\Delta_j$ and $\varepsilon^{\prime}$ into $\alpha_j$ gives $\alpha_j = \lambda R_j + \eta$. The equation $R_j = T_j + \alpha_j$ therefore becomes $(1 - \lambda)R_j = T_j + \eta$, which proves the remaining identities.
\end{proof}
\paragraph{The mechanism.}
Each threshold instance halts after its first positive answer. At every time $i$, the wrapper releases the largest threshold whose instance has fired, or the lowest threshold $T_0=1$ if no instance has fired. We denote this released value by $\bar{Y}_i$.
\begin{algorithm}[H]
\caption{\textsc{PrivateStream}}
\label{alg:privatestream}
\begin{algorithmic}[1]
\Require Base algorithm $\mathcal A$ for $F$, sensitivity bound $\Delta_F$, parameters $\varepsilon,\delta,\gamma,\kappa$, stream length $m$, and range bound $M$
\Ensure Released Stream $\bar{Y}_1,\bar{Y}_2,\dots,\bar{Y}_m$
\State $b\gets\gamma/2$, $L\gets\lceil\log_{1+b}M\rceil$, $N\gets L+1$
\State $H\gets\log(2(m+1)N/\delta)$
\State $\tau\gets\min\{\gamma/16,\varepsilon\gamma/(96NH)\}$
\State $\lambda\gets 12\tau NH/((1+\tau)\varepsilon)$
\State $\eta\gets \frac{6NH}{\varepsilon}\left((1-\tau)\Delta_F+\frac{2\kappa}{1+\tau}\right)$
\State $s\gets-1$
\For{$j=0,\dots,L$}
    \State $T_j\gets(1+b)^j$,  $R_j\gets(T_j+\eta)/(1-\lambda)$
    \State $\Delta_j\gets\frac{2\tau}{1+\tau}R_j +(1-\tau)\Delta_F+\frac{2\kappa}{1+\tau}$
    \State Initialize 
$\textsc{AboveThreshold}_j$ with threshold $T_j$, privacy $\varepsilon/N$, and sensitivity $\Delta_j$ 
\EndFor
\State Initialize $\mathcal{A}(\cdot,\tau,\kappa,\delta/2)$
\For{$i = 1, \cdots, m$}
    \State Feed $e_i$ to $\mathcal{A}$ and obtain $\widehat{F}_i$
    \For{$j=0,\dots,L$}
        \If{$\textsc{AboveThreshold}_j$ has not fired}
            \State Submit $q_{i,j} \gets \min\{\widehat{F}_i, R_j\}$ to $\textsc{AboveThreshold}_j$
            \If{$\textsc{AboveThreshold}_j$ fires}
            \State $s \gets \max\{s,j\}$
            \EndIf
        \EndIf
    \EndFor
    \State Output $\bar{Y}_i\gets 1$ if $s=-1$, and $\bar{Y}_i \gets T_s$ otherwise
\EndFor
\end{algorithmic}
\end{algorithm}
Because the threshold ladder begins at $T_0=1$, the mechanism also releases $1$ before any level fires. When the true value is below $1$ (in particular, when it is $0$), this contributes at most one unit of additive error; and the final $+2$ term in the accuracy theorem absorbs this edge case.

\subsection{Privacy}
\label{sec:privacy}
We now move to demonstrating our mechanism is private. We first show that clipping controls sensitivity whenever the base algorithm is accurate on both neighboring streams under the same random seed.

\begin{lemma} [Clipped Sensitivity]
\label{lem:sensitivity}
Fix neighboring streams $D \sim D'$, a time $i$, a threshold level $j$, and the choice of the base algorithm's random coins such that $\mathcal{E}_{\mathcal{A}}(D:r)$ and $\mathcal{E}_{\mathcal{A}}(D':r)$ both hold. Then the clipped function,
$q_{i,j}(D)$
has sensitivity bounded by $|q_{i,j}(D) - q_{i,j}(D')| \leq \Delta_j.$
    
\end{lemma}
\begin{proof}
By symmetry, assume $q_{i,j}(D) \geq q_{i,j}(D').$
If $q_{i,j}(D') = R_j$, then $q_{i,j}(D) = R_j$ and the claim is immediate. Otherwise $q_{i,j}(D') = \widehat{F}_i(D') < R_j$. The lower approximation guarantee and sensitivity of $F$ give 
$$q_{i,j}(D') \geq (1 - \tau)F(D') - \kappa \geq (1 - \tau) (F_i(D) - \Delta_F) - \kappa$$
Moreover, \(\widehat F_i(D)\geq q_{i,j}(D)\), and hence $F_i(D) \geq \frac{q_{i,j}(D)-\kappa}{1+\tau}.$
Substituting this bound yields
\begin{align*}
q_{i,j}(D)-q_{i,j}(D')
&\leq
q_{i,j}(D)
-(1-\tau)
\left(
\frac{q_{i,j}(D)-\kappa}{1+\tau}-\Delta_F
\right)
+\kappa\\
&=
\frac{2\tau}{1+\tau}q_{i,j}(D)
+(1-\tau)\Delta_F
+\frac{2\kappa}{1+\tau}\\
&\leq
\frac{2\tau}{1+\tau}R_j
+(1-\tau)\Delta_F
+\frac{2\kappa}{1+\tau} = \Delta_j.
\end{align*}
\end{proof}
We now prove that \textsc{PrivateStream} satisfies the desired privacy guarantee. The main issue is that the sensitivity bound for each instance of \textsc{AboveThreshold} holds only on the approximation event for the underlying streaming algorithm. We therefore begin with a general lemma showing how privacy under a high-probability good event converts into approximate differential privacy.

Let $r$ denote the random coins of $\mathcal{A}$. Once $r$ is fixed, for every level $j$ the full clipped query sequence $q_{1,j}(D;r),\ldots,q_{m,j}(D;r)$ is a deterministic function of the input stream. Algorithm~\ref{alg:privatestream} submits these queries to every level until that level fires, exactly as in a standard \textsc{AboveThreshold} execution; whether some other level has already fired does not change the query sequence seen by level $j$. For a stream $D$, define

$G_D := \{r : \mathcal{E}_{\mathcal{A}}(D;r) \text{ holds} \}$
and for neighboring streams $D, D'$, define
$G_{D,D^{\prime}} := G_D \cap G_{D^{\prime}}.$
The next lemma converts privacy for good random seeds into approximate privacy of the full randomized mechanism.
\begin{lemma}[Privacy lifting from good random coins]
\label{lem:privacy-lifting}
Let $r$ be input-independent random coins and let $z$ denote all remaining randomness of a mechanism $\mathcal{M}$. Suppose that, for every neighboring pair $D \sim D'$, there is a set $G_{D,D'}$ such that $\Pr_r[r\notin G_{D,D'}]\leq \delta^{\prime},$
and, for every $r\in G_{D,D'}$ and every measurable output event $S$,
$$\Pr_z[\mathcal{M}(D;r,z)\in S] \leq \exp(\varepsilon) \Pr_z[\mathcal{M}(D';r,z)\in S].$$
Then $\mathcal M$ is $(\varepsilon,\delta^{\prime})$-differentially private.
\end{lemma}
\begin{proof}
Fix neighboring streams $D\sim D'$ and an output event $S$. Then
\begin{align*}
\Pr[\mathcal{M}(D)\in S]
&=\mathbb E_r\left[\Pr_z[\mathcal{M}(D;r,z)\in S]\right]\\
&\leq
\mathbb E_r\left[
\mathbf 1_{\{r\in G_{D,D'}\}}
\Pr_z[\mathcal{M}(D;r,z)\in S]
\right]
+
\Pr_r[r\notin G_{D,D'}]\\
&\leq
\exp(\varepsilon)
\mathbb E_r\left[
\mathbf 1_{\{r\in G_{D,D'}\}}
\Pr_z[\mathcal{M}(D';r,z)\in S]
\right]
+
\delta^{\prime}\\
&\leq
\exp(\varepsilon)\Pr[\mathcal{M}(D')\in S]+\delta^{\prime}.
\end{align*}
\end{proof}

We next bound the probability of a set of random coins failing the good event for one of the neighboring streams, $D, D'$. We prove that if the probability of failing for stream $D$ is low, then the probability of failing for either $D$ or $D'$ is likewise low.
\begin{lemma}
\label{lem:pairwise-good-seed}
If, for every stream $D$, $\Pr_r[r\notin G_D]\leq \delta_A,$
then, for every neighboring pair $D\sim D'$,
$\Pr_r[r\notin G_{D,D'}]\leq 2\delta_A.$
\end{lemma}

\begin{proof}
$G_{D,D'}=G_D\cap G_{D'}$, so a union bound gives
$\Pr_r[r\notin G_{D,D'}]
\leq
\Pr_r[r\notin G_D] + \Pr_r[r\notin G_{D'}]
\leq 2\delta_A.
$
\end{proof}
The privacy of \textsc{AboveThreshold} for queries that are deterministic and guaranteed to satisfy a given sensitivity is standard, c.f., \cite{dwork2014algorithmic}. 
We provide a full analysis with proof in \Cref{lem:privacy:svt-standard}. It is a necessary stepping stone to guaranteeing privacy in the case where query inputs $\widehat{F_i}$ may be nondeterministic and are not assumed to meet sensitivity. We apply this lemma to demonstrate that each instance of \textsc{AboveThreshold} is $\varepsilon^{\prime}$-differentially private when conditioned on the good event.

\begin{corollary}[Privacy of an SVT instance]
\label{cor:privacy:svt}
Fix a threshold $j$, neighboring streams $D, D'$, and a set of coins $r \in G_{D,D'}.$ Then for every output event $S_j$
$$\Pr_{z_j}[\operatorname{SVT}_j(D;r,z_j)\in S_j] \leq \exp(\varepsilon') \Pr_{z_j}[\operatorname{SVT}_j(D';r,z_j)\in S_j]. 
$$
\end{corollary}

\begin{proof}
For fixed $r\in G_{D,D'}$, every clipped query is deterministic, and \Cref{lem:sensitivity} bounds the difference between $D$ and $D'$ by $\Delta_j$. The claim follows from \Cref{lem:privacy:svt-standard} with sensitivity parameter $\Delta_j$ and privacy parameter $\varepsilon'$.
\end{proof}
As \textsc{PrivateStream} consists of multiple instances of \textsc{AboveThreshold} running in parallel, the privacy guarantee of the private streaming algorithm naturally follows from general privacy composition.
\begin{theorem} [Privacy of \textsc{PrivateStream}] \label{thm:privacy:stream}
Suppose the base algorithm is run with failure probability $\delta/2$ and suppose $\lambda < 1$. Then \textsc{PrivateStream} is $(\varepsilon, \delta)$-differentially private. 
\end{theorem}
\begin{proof}
    Fix neighboring streams $D\sim D'$ and a set of coins $r\in G_{D,D'}$.  Recall from \Cref{cor:privacy:svt}, that each of the $N$ threshold instances is $\varepsilon'=\varepsilon/N$-DP for a fixed $r$. There are at most $N$ such instances, thus basic composition gives $\varepsilon$-privacy conditioned on $r$. By \Cref{lem:pairwise-good-seed}, $\Pr_r[r\notin G_{D,D'}]\leq 2\delta_A.$ Thus, since the failure probability of $\mathcal{A}$ is $\delta/2$, applying \Cref{lem:privacy-lifting} gives $(\varepsilon,\delta)$-differential privacy. Thus the claimed privacy guarantee follows. 
\end{proof}
\subsection{Accuracy}
\label{sec:accuracy}
We now move to proving the claimed accuracy guarantees. Let $\mathcal{E}_{\mathcal{A}}$ be the event that the outputs of $\mathcal{A}$ are $(\tau,\kappa)$-accurate at each time. For each threshold level $j$, let $\mathcal{E}_j$ be the event that the noisy threshold differs from $T_j$ by at most $2H \Delta_j/\varepsilon^{\prime}$ and every query noise has magnitude at most $4H \Delta_j/\varepsilon^{\prime}$. This is exactly the event of \Cref{lem:AT-accuracy}, which bounds the failure probability with $\Pr[\mathcal{E}_j^c]\leq(m+1)e^{-H}=\frac{\delta}{2N}.$
This result for \textsc{AboveThreshold} is standard~ \cite{dwork2014algorithmic}. 
Let $\mathcal{E_\mathrm{AT}} := \bigcap_{j=0}^{L} \mathcal{E}_j.$ Since the base failure probability is $\delta/2$,
$\Pr[\mathcal{E}_\mathcal{A}(D) \cap \mathcal{E}_\mathrm{AT}] \geq 1 - \delta.$ We later condition on the joint good event
$\mathcal{E}_{\mathcal{A}}\cap\mathcal{E}_{\mathrm{AT}}$. We first translate whether a threshold has fired into corresponding upper and
lower bounds on the true value $F_i(D)$.
\begin{lemma} [Threshold Crossing Bounds] \label{lem:accuracy:threshold}
Assume the event $\mathcal{E}_{\mathcal{A}} \cap \mathcal{E}_j$ holds. Then:

\begin{enumerate}
    \item if level $j$ has not fired by time $i$, 
    $F_i(D) \leq \frac{T_j + \alpha_j + \kappa}{1 - \tau};$
    \item If level $j$ fires at time $k$,
    $F_k(D) \geq \frac{T_j - \alpha_j - \kappa}{1 + \tau}.$
\end{enumerate}
\end{lemma}
\begin{proof}
First, suppose level $j$ has not fired by time $i$. Then on $\mathcal{E}_j$, the \textsc{AboveThreshold} accuracy guarantee applies to the clipped query and gives $q_{i,j}(D)<T_j+\alpha_j.$ By \Cref{prop:optimal-clipping}, $T_j + \alpha_j = R_j$. Hence, $q_{i,j} < R_j.$ Since $q_{i,j}(D) = \min\{\widehat{F}_i(D), R_j\}$, clipping cannot be active and therefore $q_{i,j}(D) = \widehat{F}_i(D)$. Using the lower tracker guarantee,
$\widehat{F}_i(D) \geq (1 - \tau)F_i(D) - \kappa,$
we obtain $F_i(D) \leq \frac{\widehat{F}_i(D) + \kappa}{1 - \tau} \leq \frac{T_j + \alpha_j + \kappa}{1 - \tau}.$

Now, instead suppose level $j$ has output $\top$ at time $k$. On $\mathcal{E}_j$, $q_{k,j}(D) \geq T_j - \alpha_j$. Because $q_{k,j}(D)\leq\widehat F_k(D)$, we also have $\widehat F_k(D)\geq T_j-\alpha_j$. Combining this with
$\widehat{F}_k(D) \leq (1 + \tau)F_k(D) + \kappa,$
we obtain 
$F_k(D) \geq \frac{\widehat{F}_k(D) - \kappa}{1 + \tau} \geq \frac{T_j - \alpha_j - \kappa}{1 + \tau}.$
\end{proof}
At any time, the largest fired level gives a lower certificate on $F_i(D)$,
while the next unfired level gives an upper certificate. Combining these two
bounds sandwiches the released value $\bar{Y}_i$ around $F_i(D)$.
\begin{theorem} [Accuracy on the Good Event]
\label{thm:accuracy-good-event}
Assume $\mathcal{E}_{\mathcal{A}}\cap \mathcal{E}_{\mathrm{AT}}$ holds. Then for every time $i$ the following properties hold:
$$\bar{Y}_i \leq \frac{(1+\tau)(1-\lambda)}{1-2\lambda}F_i(D) + \frac{\eta+(1-\lambda)\kappa}{1-2\lambda}
+1,$$
$$F_i(D) \leq \frac{1+b}{(1-\lambda)(1- \tau)}\bar{Y}_i + \frac{1 + \eta + (1-\lambda)\kappa}{(1-\lambda)(1 - \tau)}.$$
\end{theorem}
\begin{proof}
Fix a time $i$ and let $s$ be the largest level that has fired by time $i$, with $s = -1$ if none have fired. 

We first upper bound the released value $\bar{Y}_i$. If $s=-1$, then
$\bar{Y}_i=1$, and the claim follows from the nonnegativity of
$F_i(D)$, $\eta$, and $\kappa$. Otherwise, level $s$ fired at some time $k \leq i$. By monotonicity and \Cref{lem:accuracy:threshold},
$T_s \leq (1 + \tau)F_i(D) + \alpha_s + \kappa.$

By \Cref{prop:optimal-clipping}, $\alpha_s = (\lambda T_s + \eta)/(1-\lambda)$. Rearranging gives the first statement. We next upper bound $F_i(D)$ in terms of $\bar{Y}_i$. If $s = L$, then $F_i(D) \leq M \leq T_L = \bar{Y}_i$, so the claim is immediate. Otherwise, level $s+1$ has not fired by time $i$. Using $\alpha_{s+1} =  (\lambda T_{s+1} + \eta)/(1-\lambda)$ and \Cref{lem:accuracy:threshold} gives

$$F_i(D) \leq \frac{T_{s+1} + \eta + (1- \lambda)\kappa}{(1 -\lambda)(1 - \tau)}.$$

If $s \geq 0$, $T_{s+1} = (1 + b)T_s = (1+b)\bar{Y}_i$. If $s = -1$, then $T_{s+1} = T_0 = 1 .$ In both cases, $T_{s+1} \leq (1+b)\bar{Y}_i + 1$ which yields the second statement.
\end{proof}
It remains to choose the internal parameters so that both multiplicative
factors above are at most $1+\gamma$ and the remaining error is additive.
\begin{corollary}[Tuned Accuracy Guarantee]
\label{cor:accuracy:tuned}
Fix $0<\gamma\leq1$ and $0<\delta<1$, and choose
$b=\frac{\gamma}{2}$ and $
\tau=\min\left\{
\frac{\gamma}{16},
\frac{\varepsilon\gamma}{96NH}
\right\}.$
Then $\lambda\leq\gamma/8$, and \textsc{PrivateStream} is
$$
\left(
\gamma,
\frac{12NH}{\varepsilon}(\Delta_F+2\kappa)+2\kappa+2, \delta
\right)
\text{-accurate}.
$$
\end{corollary}
\begin{proof}
The choice of $\tau$ gives
$\lambda = \frac{12\tau NH}{(1 + \tau)\varepsilon}\leq\frac{\gamma}{8}.$
For $0 \leq \gamma \leq 1$,
$$\frac{(1+\tau)(1-\lambda)}{1-2\lambda}
\leq
\frac{(1+\gamma/16)(1-\gamma/8)}{1-\gamma/4}
\leq1+\gamma,$$
and
$$\frac{1+b}{(1-\lambda)(1-\tau)}
\leq
\frac{1+\gamma/2}{(1-\gamma/8)(1-\gamma/16)}
\leq1+\gamma.$$
Moreover, $\lambda\leq1/8$ and $\tau\leq1/16$, so both additive terms in \Cref{thm:accuracy-good-event} are at most $2\eta + 2\kappa + 2$. Since $\eta \leq \frac{6NH}{\varepsilon} (\Delta_F + 2\kappa)$ the stated additive error follows.  Rearranging the second bound and using
$1/(1+\gamma)\geq1-\gamma$ gives the corresponding lower bound on
$\bar{Y}_i$.

Finally, the base approximation event fails with probability at most $\delta/2$, and a union bound over the at most $N$ threshold events adds at most $N\delta/(2N)=\delta/2$. Thus the joint good event holds with probability at least $1-\delta$.
\end{proof}
\subsection{Main Result}
Combining the privacy guarantee with the tuned accuracy bound yields the main
transformation.
\begin{theorem}
\label{thm:main}
Let $F$ be a nonnegative, monotone, insertion-only, polynomially bounded, streaming statistic with
sensitivity $\Delta_F$ and range bounded by $M\geq 1$. Suppose there exists a
tunable continual approximation algorithm $\mathcal A$ such that, when run
with parameters $(\alpha,\kappa,\delta)$, it is
$(\alpha,\kappa,\delta)$-continually accurate and uses
$S_{\mathcal A}(m,\alpha,\kappa,\delta)$ space. Fix desired parameters $\gamma\in(0,1],\varepsilon\in(0,1],\delta\in(0,1),$ and $\kappa\geq 0$.

Run \textsc{PrivateStream} with threshold spacing $b=\gamma/2$, additive parameter $\kappa$, and top-level failure probability $\delta$ letting the wrapper
choose its internal multiplicative parameter $\tau_{\mathrm{alg}}$ according
to Algorithm~\ref{alg:privatestream}. For the stated space bound, define the conservative parameter
$$
\tau_{\mathrm{wc}}
:=
\frac{\varepsilon\gamma^2}
{288(\log{M} + 1)\log\!\left(\frac{6(m+1)(\log{M}+1)}{\gamma\delta}\right)},
$$
Then the resulting continual release mechanism satisfies all of the following.

\begin{enumerate}
    \item \textbf{Privacy.} The complete released stream is
    $(\varepsilon,\delta)$-differentially private.

    \item \textbf{Accuracy.} With probability at least $1-\delta$,
    simultaneously for every time $i$,
    $$
    (1-\gamma)F_i(D)-\rho
    \leq
    \bar{Y}_i
    \leq
    (1+\gamma)F_i(D)+\rho,
    $$
    where
    $$
    \rho
    :=
    \frac{36(\Delta_F+2\kappa)}{\varepsilon\gamma}
    (\log{M} + 1)
    \log\left(\frac{6(m+1)(\log{M}+1)}{\gamma\delta}\right) + 2\kappa + 2
    $$
    Equivalently, \textsc{PrivateStream} is
    $(\gamma,\rho,\delta)$-accurate at all times in the data stream. 
    
    \item \textbf{Space.} The mechanism uses space at most
    $$
    S_{\mathcal A}\left(
        m,
        \frac{\varepsilon\gamma^2}
        {288(\log{M} + 1)\log\left(\frac{6(m+1)(\log{M}+1)}{\gamma\delta}\right)},
        \kappa,
        \frac{\delta}{2}
    \right)
    +
    O\left(\frac{\log{M} + 1}{\gamma}\right).
    $$
\end{enumerate}
\end{theorem}

\begin{proof}
Let
$$
N=\ceil{\log_{1+\gamma/2}M}+1,
\qquad
Q=\log\!\left(\frac{6(m+1)(\log M+1)}{\gamma\delta}\right)
$$
where $N$ is the number of threshold levels and $Q$ is a convenient upper bound on the Laplace tail parameter. For
$x\geq 0$, $\log(1+x)\geq x/(1+x)$. Since $0<\gamma\leq 1$, this gives
$
\log(1+\gamma/2)
\geq
\frac{\gamma}{2+\gamma}
\geq
\frac{\gamma}{3}.
$
Consequently,
$
N
\leq
\frac{3(\log M + 1)}{\gamma}.
$
Therefore
$H = \log\!\left( \frac{2(m+1)N}{\delta}\right) \leq Q,$
so
$$NH \leq \frac{3(\log M+1)Q}{\gamma}\qquad\text{and}\qquad\frac{\varepsilon\gamma}{96NH}
\geq
\frac{\varepsilon\gamma^2}{288(\log M+1)Q}
=
\tau_{\mathrm{wc}}.$$
Moreover, since $\varepsilon,\gamma\leq1$ and $Q>1$, we have $\tau_{\mathrm{wc}}\leq\gamma/16$. Hence
$\tau_{\mathrm{alg}}= \min\left\{\frac{\gamma}{16},\frac{\varepsilon\gamma}{96NH}\right\} \geq \tau_{\mathrm{wc}}, $
and \Cref{cor:accuracy:tuned} applies to the parameter $\tau_{\mathrm{alg}}$ chosen by the algorithm. Its additive error is at most
$\frac{12NH}{\varepsilon}(\Delta_F+2\kappa)+2\kappa+2,$
which is bounded by the stated value of $\rho$. The mechanism stores one copy of $\mathcal A$, one constant-size state
for each of the $N$ \textsc{AboveThreshold} instances, and the largest fired
level.  It therefore uses
$$S_{\mathcal A}\!\left(m,\tau_{\mathrm{alg}},\kappa,\frac{\delta}{2}\right)+O(N)
=
S_{\mathcal A}\!\left(m,\tau_{\mathrm{alg}},\kappa,\frac{\delta}{2}\right)
+O\!\left(\frac{\log M+1}{\gamma}\right)$$
space.
Since $\tau_{\mathrm{alg}} \geq \tau_{\mathrm{wc}}$ we can use the monotone space convention from the preliminaries. This yields exactly the
space bound stated. Lastly, privacy follows from \Cref{thm:privacy:stream}.
\end{proof}

When $M\leq m^c$, the range-dependent bounds in \Cref{thm:main}
satisfy $\log M=O(\log m)$, giving $N=O(\log m/\gamma)$.
For fixed privacy and accuracy parameters, the wrapper adds only
$O(\log m)$ space and requires inverse-polylogarithmic base accuracy.
Thus, for base algorithms whose space is polynomial in $1/\tau$,
privatization incurs only a polylogarithmic space overhead.

\section{Applications}
\label{sec:applications-main}

We illustrate the framework briefly through three representative
streaming problems: distinct elements, $L_p$ norm estimation
with fixed $p\in[1,2]$, and longest increasing subsequence.
We summarize their guarantees here and discuss their
relationship to existing private continual-release algorithms.
The full comparison tables in \Cref{app:comparisons}
compare privacy, multiplicative and additive error,
and space requirements with both non-private base algorithms
and application-specific private algorithms.
These comparisons highlight the tradeoffs between our
reusable black-box transformation and specialized methods,
some of which provide pure privacy or support deletions.
Complete application statements and proofs appear in
Appendix~\ref{app:applications}, including additional
ranges of $p$, regression, low-rank approximation,
clustering, and union volume estimation.
The pure-private conversion and its additional costs
are analyzed in Appendix~\ref{sec:pure-dp}.
The examples presented here cover both frequency statistics
and a sequence-dependent optimization problem, demonstrating
that the transformation does not rely on a particular
sketching method or data representation.

All three statistics satisfy the same structural conditions.
They are nonnegative and monotone under insertions, have
sensitivity at most one, and take values at most $m$.
For distinct elements, inserting one event introduces at
most one new element. For $L_p$ norms with $p\geq1$, the
sensitivity bound follows from the reverse triangle
inequality. For longest increasing subsequence, inserting
or deleting one event changes the optimum by at most one.
Consequently, the three instantiations share the same
additive-error guarantee; their different space bounds
come from the underlying approximation algorithms.

Applying \Cref{thm:main} to the algorithms of
\cite{10.1145/3309193},
\cite{blasiok_et_al:LIPIcs.APPROX-RANDOM.2017.32}, and
\cite{GopalanJayramKrauthgamerKumar2007}, respectively,
gives the following consequences.

\begin{corollary}
\label{cor:applications-summary}
Fix $\varepsilon,\gamma\in(0,1]$ and $\delta\in(0,1)$.
For each statistic in \Cref{tab:applications-summary},
there is an $(\varepsilon,\delta)$-differentially private
continual-release algorithm that is
$(\gamma,\rho,\delta)$-continually accurate, where
\[
\rho =
O\left(
  \frac{\log(m+1)}{\varepsilon\gamma}
  \log\frac{m+1}{\gamma\delta}
  +1
\right).
\]
For distinct elements and frequency norms, assume the
universe size satisfies $n=\operatorname{poly}(m)$.
\end{corollary}

\begin{table}[H]
\centering
\caption{Space bounds for the approximate-private
instantiations. All bounds count words under the space
model in \Cref{sec:preliminaries}.
The $\widetilde O$ notation suppresses logarithmic
factors in $m,n,1/\varepsilon,1/\gamma$, and $1/\delta$.}
\label{tab:applications-summary}
\begin{tabular}{ll}
\toprule
Statistic & Space \\
\midrule
Distinct elements
& $\widetilde O\left(\frac{1}{\varepsilon^2\gamma^4}\right)$
\\
$L_p$ norm, fixed $p\in[1,2]$
& $\widetilde O\left(\frac{1}{\varepsilon^2\gamma^4}\right)$
\\
Longest increasing subsequence
& $\widetilde O\left(\frac{\sqrt m}{\sqrt\varepsilon\,\gamma}\right)$
\\
\bottomrule
\end{tabular}
\end{table}

The corollary follows by setting $\Delta_F=1$, $M=m$,
and $\kappa=0$ in \Cref{thm:main} and substituting the
corresponding base-algorithm space bounds.
The framework runs each base algorithm at internal
multiplicative accuracy
$\tau=\Theta(\varepsilon\gamma^2/
(\log(m+1)\log((m+1)/(\gamma\delta))))$,
up to absolute constants.
For fixed privacy and accuracy parameters, this requires
only a polylogarithmic tightening of the base accuracy.
Thus the distinct-elements and frequency-norm
instantiations retain polylogarithmic space, while the
longest-increasing-subsequence instantiation retains
sublinear space with a $\sqrt m$ dependence.

The common accuracy guarantee separates two sources of
error. The parameter $\gamma$ controls error proportional
to the statistic, while $\rho$ accounts for the threshold
noise and the behavior at small values.
When $F_i(D)$ is large compared with $\rho/\gamma$,
the guarantee is predominantly multiplicative.
At smaller values, the additive term can dominate.
Importantly, the guarantee holds simultaneously at every
time, and privacy protects the complete released
transcript.

These results complement application-specific private
streaming algorithms. Epasto et al.~\cite{epasto2023differentially}
give pure-private continual-release algorithms for frequency
statistics. Aamand et al.~\cite{aamand2026skirting} also
support turnstile streams, where deletions can make the
statistic decrease and our monotonicity-based analysis
does not directly apply. The framework therefore does
not uniformly improve the best specialized bounds;
its benefit is that the same mechanism and privacy
analysis can be reused across different statistics.

For longest increasing subsequence, the comparison
highlights a different tradeoff. The guarantee of
\cite{bonomi2016lis} has an $O(\log m)$ multiplicative
factor and linear space, whereas our approximate-private
instantiation provides tunable multiplicative error
with sublinear space for fixed parameters.
This illustrates how an existing non-private approximation
algorithm can yield a private continual-release guarantee
without a new problem-specific privacy mechanism.

\clearpage

\def\shortbib{0}
\bibliographystyle{alpha}
\bibliography{references}
\clearpage
\appendix

\section{Approximate DP to Pure DP}
\label{sec:pure-dp}
The approximate privacy loss in \Cref{thm:main} arises from the failure event of the randomized base algorithm. If the base algorithm is deterministic, then the privacy analysis already gives pure $\varepsilon$-differential privacy. For a randomized algorithm, we obtain a generic pure private variant by exploiting the finite output range of \textsc{PrivateStream}. 
The construction runs the approximate-private mechanism on most executions and, with a small data-independent probability, replaces the full output stream with a uniformly random valid threshold output stream. The random branch places positive probability on every possible transcript and thereby absorbs the additive privacy loss. We use the following folklore result to justify the $\varepsilon$-DP claim of \textsc{PurePrivateStream}. We include its proof for completeness.
\begin{theorem}[Approximate DP to Pure DP]
\label{thm:approxtopure}
Let \(\mathcal{A'} : \mathcal{D} \to \mathcal{R}\) be an \((\varepsilon,\delta)\)-differentially private algorithm. If
$$\delta \leq \frac{(e^{\varepsilon}-1)p}{|\mathcal{R}|(1-p)},$$
then there exists an \(\varepsilon\)-differentially private algorithm \(\mathcal{A''}\) defined by

$$\mathcal{A}''(D)=
\begin{cases}
\mathcal{A}'(D), & \text{with probability } 1-p,\\[1ex]
\operatorname{random}(\mathcal{R}), & \text{with probability } p,
\end{cases}$$
where \(\operatorname{random}(\mathcal{R})\) denotes an element chosen uniformly at random from the range \(\mathcal{R}\).

Moreover, if $\mathcal A'$ satisfies an accuracy guarantee with failure probability at most $\beta$, and every execution of the uniform branch is conservatively counted as a failure, then $\mathcal A''$ satisfies the same accuracy guarantee with failure probability at most
$$p+(1-p)\beta.$$
\end{theorem}
\begin{proof}
    The proof is the same argument used in \cite{blocki2023makeapproximationalgorithmprivate}. Assume $\mathcal{A}'$ is $(\varepsilon,\delta)$-DP and that $\delta \leq \frac{(e^{\varepsilon}-1)p}{|\mathcal{R}|(1-p)}$ where $p$ is fixed satisfying $0\leq p<1$. Let $\mathcal{A}''$ be as defined above, $D\sim D'$ be neighboring streams, and $y\in \mathcal{R}$ be arbitrary. Then using the $(\varepsilon,\delta)$-DP guarantee of $\mathcal{A}'$ and the definition of $\mathcal{A}''$,
    \begin{align*}
        \Pr[\mathcal{A}''(D)=y]=&\Pr[\mathcal{A}'(D)=y](1-p)+\frac{p}{|\mathcal{R}|}\\
        \leq&(1-p)(e^\varepsilon\Pr[\mathcal{A}'(D')=y]+\delta)+\frac{p}{|\mathcal{R}|}\\
        =&(1-p)(e^\varepsilon\cdot \frac{\Pr[\mathcal{A}''(D')=y]-\frac{p}{|\mathcal{R}|}}{1-p}+\delta)+\frac{p}{|\mathcal{R}|}.
    \end{align*}
    We cancel out terms in the last expression in the above string of inequalities, then use the assumed bound on $\delta$ to obtain the desired pure privacy bound,
    \begin{align*}
        \Pr[\mathcal{A}''(D)=y]\leq&e^\varepsilon\Pr[\mathcal{A}''(D')=y]+(1-p)\delta+(1-e^\varepsilon)\cdot\frac{p}{|\mathcal{R}|}\\
        \leq&e^\varepsilon\Pr[\mathcal{A}''(D')=y]+(e^{\varepsilon}-1)\cdot\frac{p}{|\mathcal{R}|}+(1-e^\varepsilon)\frac{p}{|\mathcal{R}|}\\
        =&e^\varepsilon\Pr[\mathcal{A}''(D')=y]
    \end{align*}
    The accuracy claim follows because the mechanism succeeds whenever it selects the $\mathcal{A}'$ branch and that branch is accurate, an event of probability at least $(1-p)(1-\beta)$.
\end{proof}
In order to utilize \Cref{thm:approxtopure}, we must establish the cardinality of our range of outputs, $|\mathcal{R}|$. To do this, recall that \textsc{PrivateStream} releases $\bar{Y}_i \in \{T_0, \dots, T_{N-1}\}$ at every time $i$. Since levels fire monotonically, the complete released output stream belongs to 
$$\mathcal{R}=\{(T_{i_1},T_{i_2},\dots,T_{i_m}) : \forall j, 0\leq i_j\leq i_{j+1}\leq N - 1\},$$ 
Thus a standard stars-and-bars count nets the following.
\begin{fact}
\label{fact:output-card}
The range of outputs of \textsc{PrivateStream} has cardinality $$|\mathcal{R}|=\binom{N+m-1}{m},$$ for stream length $m$ and $N$ SVT thresholds.
\end{fact}
Before running  \textsc{PrivateStream}, we will flip a coin, which may result in random outputs at each time. To sample uniformly from $\mathcal R$ online, we use a stars-and-bars representation. A weakly increasing sequence of $m$ indices from $\{0, \cdots, N-1\}$ is in bijection with a string containing $m$ stars and $N-1$ bars; the index of the $i$th output is the number of bars appearing before the $i$th star. A uniformly random string can be generated sequentially. If $a$ stars and $c$ bars remain, the next symbol is a star with probability $\frac{a}{a + c}$ and a bar with probability $\frac{c}{a + c}$.

\begin{algorithm}
\caption{\textsc{PurePrivateStream}}
\label{alg:pureprivatestream}
\begin{algorithmic}[1]
\Require A configured instance of \textsc{PrivateStream}, number of levels $N$, stream length $m$, and mixing probability $p$
\Ensure Released stream $Z_1,Z_2,\dots,Z_m$
\State Sample $X\gets \operatorname{Bern}(p)$
\If{$X=0$}
    \For{$i=1,\dots,m$}
        \State Feed $e_i$ to \textsc{PrivateStream} and release its output as $Z_i$
    \EndFor
\Else
    \State $a \gets m, c \gets N - 1, t \gets 0$    \For{$i=1,\dots,m$}
        \While{$c > 0$}
            \State Sample $B\gets \operatorname{Bern}(\frac{a}{a+c})$
            \If{$B = 1$}
                \State \textbf{break}
            \Else
                \State $c \gets c - 1, t \gets t + 1$
            \EndIf
        \EndWhile
    \State Output $Z_i\gets T_t$
    \State $a \gets a - 1$
    \EndFor
\EndIf
\end{algorithmic}
\end{algorithm}

\begin{lemma}
\label{lem:privacy:random}
    With probability $p$, \textsc{PurePrivateStream} outputs $\operatorname{Random}(\mathcal{R})$, otherwise it outputs a series of approximations identical to \textsc{PrivateStream}.
\end{lemma}
\begin{proof}
Condition on $X = 1$. The random branch generates a string with $m$ stars and $N-1$ bars and outputs $T_t$ at each star, where $t$ is the number of bars seen so far. This is the stars-and-bars bijection with $\mathcal R$.

Fix any particular ordering of the $m$ stars and $N-1$ bars. During the sequential sampler, the numerator contributed by all star choices is $m!$, the numerator contributed by all bar choices is $(N-1)!$, and the product of the denominators is $(m+N-1)!$. Hence this ordering is generated with probability
$$ \frac{m!(N-1)!}{(m+N-1)!} = \frac{1}{\binom{N+m-1}{m}}
= \frac{1}{|\mathcal R|}.
$$
Thus the random branch is uniform on $\mathcal R$. The $X=0$ branch runs \textsc{PrivateStream} unchanged.
\end{proof}

This establishes a pure privacy guarantee for the modified algorithm. Our framework provides a $(\varepsilon,\delta)$-differentially private stream by running the \textsc{PrivateStream} wrapper, and a $\varepsilon$-differentially private stream by running the \textsc{PurePrivateStream} wrapper. This privacy guarantee comes at the expense of the approximation failure parameter, however the parameter remains tunable.
\begin{theorem}[Guarantee for \textsc{PurePrivateStream}]
\label{thm:main:pure}
Let $F$ be a nonnegative, monotone, insertion-only, polynomial bounded, streaming statistic with
sensitivity $\Delta_F$ and range bounded by $M\geq 1$. Suppose there exists a
tunable continual approximation algorithm $\mathcal A$ such that, when run
with parameters $(\alpha,\kappa,\delta)$, it is
$(\alpha,\kappa,\delta)$-continually accurate and uses
$S_{\mathcal A}(m,\alpha,\kappa,\delta)$ space. Fix desired parameters $\gamma\in(0,1],\varepsilon\in(0,1],\delta\in(0,1),$ and $\kappa\geq 0$.
Define
$$
N:=\ceil{\log_{1+\gamma/2}M}+1,
\qquad 
K:=\binom{N+m-1}{m}, \qquad \delta_*
:=
\frac{\delta(e^{\varepsilon}-1)}
{e^{\varepsilon}-1+(1-\delta)K}
$$
and
$$
p_*
:=
\frac{\delta_*K}{e^{\varepsilon}-1+\delta_*K}
=
\frac{\delta K}{e^{\varepsilon}-1+K}.
$$
$b=\gamma/2$, additive parameter $\kappa$, failure probability $\delta_*$. 
splitting the privacy and SVT failure budgets uniformly across threshold instances, and
letting the wrapper
choose its internal multiplicative parameter $\tau_{\mathrm{alg}}$ according
to Algorithm~\ref{alg:privatestream}. For the stated space bound, define the conservative parameter
$$
\tau_*
:=
\frac{\varepsilon\gamma^2}
{288(\log{M} + 1)\log\!\left(\frac{6(m+1)(\log{M}+1)}{\gamma\delta_*}\right)}.
$$ Use this mechanism as the approximate-private branch of
\textsc{PurePrivateStream} with mixing probability $p=p_*$. Then
the resulting continual release mechanism satisfies all of the
following.

\begin{enumerate}
    \item \textbf{Privacy.} The complete released stream is $\varepsilon$-differentially private.

    \item \textbf{Accuracy.} With probability at least $1-\delta$, simultaneously for every time $i$,
    $$
    (1-\gamma)F_i(D)-\rho_*
    \leq
    Z_i
    \leq
    (1+\gamma)F_i(D)+\rho_*,
    $$
    where
    $$
    \rho_*
    :=
    \frac{36(\Delta_F+2\kappa)(\log M+1)}{\varepsilon\gamma}
    \log\!\left(
    \frac{6(m+1)(\log M+1)}{\gamma\delta_*}
    \right)
    +2\kappa+2.
    $$
    Equivalently, \textsc{PrivateStream} is $(\gamma,\rho_*,\delta)$-continually accurate.

    \item \textbf{Space.} The mechanism uses space at most

    $$
    S_{\mathcal A}\Bigg(
        m,
        \frac{\varepsilon\gamma^2}
{288(\log{M} + 1)\log\!\left(\frac{6(m+1)(\log{M}+1)}{\gamma\delta_*}\right)}
        \kappa,
        \frac{\delta_*}{2}
    \Bigg)
    +
    O\left(\frac{\log(M+1)}{\gamma}\right).
    $$
\end{enumerate}
\end{theorem}

\begin{proof}
We first apply \Cref{thm:main} to the
\textsc{PrivateStream} branch with failure parameter $\delta_*$ in
place of $\delta$. Under the configuration stated in the theorem,
we get all three guarantees needed for that branch: namely
it is $(\varepsilon,\delta_*)$-differentially private, it is
$(\gamma,\rho_*,\delta_*)$-continually accurate, and uses the stated space.

We now move to analyzing the outer randomization performed by
\textsc{PurePrivateStream}. By \Cref{fact:output-card} and \Cref{lem:privacy:random}, the other branch is uniform on a range of size $K$. Moreover,
$$\frac{(e^{\varepsilon}-1)p_*}{K(1-p_*)}= \delta_*.
$$
Thus the hypothesis of \Cref{thm:approxtopure} holds with equality
for the $(\varepsilon,\delta_*)$-private
\textsc{PrivateStream} branch. Therefore,
\textsc{PurePrivateStream} is $\varepsilon$-differentially private.

For accuracy, we count every execution of the random branch as a failure. The success probability is therefore at least $(1-p_*)(1-\delta_*).$ Writing $a=e^{\varepsilon}-1$, the definitions of $p_*$ and $\delta_*$
give
$$
1-p_* = \frac{a+(1-\delta)K}{a+K}, \qquad 1-\delta_* = \frac{(1-\delta)(a+K)} {a+(1-\delta)K}.
$$
Hence,
$(1-p_*)(1-\delta_*)=1-\delta.$
Therefore the mechanism is
$(\gamma,\rho_*,\delta)$-continually accurate.

Finally, only one of the two branches is executed. The
\textsc{PrivateStream} branch has the space bound supplied by
\Cref{thm:main}, while the random branch only maintains the current
threshold index and the streaming counters. This additional
bookkeeping is absorbed by the $O(N)$ wrapper term. The claimed
space bound follows.
\end{proof}

\section{Full Applications}
\label{app:applications}
The strength of our framework is that obtaining a private continual-release algorithm no longer requires designing a new privacy mechanism for each streaming problem. Instead, each application follows the same recipe: we identify a nonnegative monotone statistic, bound its sensitivity and range, and supply an existing tunable continual approximation algorithm; \Cref{thm:main} then gives the corresponding private guarantee as a black-box consequence. The examples below show that these conditions arise across several different areas of streaming, ranging from classical frequency statistics and sequence problems to numerical linear algebra, clustering, and geometric estimation. In particular, under the standard parameter regimes considered below, these statistics have polynomially bounded range, so the transformation introduces only $O(\log m/\gamma)$ threshold instances and requires only a polylogarithmic tightening of the base algorithm's accuracy. Thus, whenever the underlying algorithm retains sublinear space under this tightening, the resulting private continual-release algorithm does as well.
\subsection{Comparisons with Existing Literature}
\label{app:comparisons}
We compare the framework's instantiations with their
non-private base algorithms in \Cref{tab:comparison}
and with application-specific private algorithms in
\Cref{tab:private-comparison}.
The pure-private guarantees use the conversion in
Appendix~\ref{sec:pure-dp}.

\begingroup
\small
\renewcommand{\arraystretch}{1.25}
\setlength{\tabcolsep}{3pt}
\begin{longtable}{@{}>{\raggedright\arraybackslash}p{\dimexpr 0.15\linewidth-2\tabcolsep\relax}>{\raggedright\arraybackslash}p{\dimexpr 0.11\linewidth-2\tabcolsep\relax}>{\raggedright\arraybackslash}p{\dimexpr 0.105\linewidth-2\tabcolsep\relax}>{\raggedright\arraybackslash}p{\dimexpr 0.25\linewidth-2\tabcolsep\relax}>{\raggedright\arraybackslash}p{\dimexpr 0.385\linewidth-2\tabcolsep\relax}@{}}
\caption{Comparison of privacy, additive error, and space complexity for various problems.}\label{tab:comparison}\\
\toprule
Application & Reference & Privacy & Additive error & Space complexity\\
\midrule
\endfirsthead
\caption[]{Comparison of privacy, additive error, and space complexity for various problems (continued).}\\
\toprule
Application & Reference & Privacy & Additive error & Space complexity\\
\midrule
\endhead
\midrule
\multicolumn{5}{r}{Continued on next page}\\
\endfoot
\bottomrule
\endlastfoot
Distinct Elements & \cite{10.1145/3309193} & N/A & $0$ & $O\Bigl(\log n+\frac1{\gamma^2}\Bigl(\log\frac1\delta+\log\log n\Bigr)\Bigr)$\\*[3pt]
 & This Work & $(\varepsilon,\delta)$ & $\widetilde O\Bigl(\frac{1}{\varepsilon\gamma}\log m\cdot\log\frac{m}{\delta}\Bigr)$ & $\begin{aligned}[t]&\textstyle \widetilde O\Bigl(\frac{1}{\varepsilon^2\gamma^4}\Bigl(\log^4m\log\frac1\delta\\&\textstyle +\log^2m\log^3\frac1\delta\Bigr)\Bigr)\end{aligned}$\\
\midrule
$L_p$ Norm Approxi\-mation, $p\in(0,1)$ & \cite{blasiok_et_al:LIPIcs.APPROX-RANDOM.2017.32} & N/A & $0$ & $O\Bigl(\frac{\log m}{\gamma^2}\Bigl(\log\log m+\log\frac1\gamma+\log\frac1\delta\Bigr)\Bigr)$\\*[3pt]
 & This Work & $(\varepsilon,\delta)$ & $\begin{aligned}[t]&\textstyle \widetilde O\Bigl(\frac{1-p}{p^2\varepsilon\delta}\cdot m^{1-p}n^{\frac{(1-p)^2}{p}}\\&\textstyle \cdot\log m\log n\cdot\log\frac m\delta\Bigr)\end{aligned}$ & $\begin{aligned}[t]&\textstyle \widetilde O\Bigl(\frac{\log m}{\varepsilon^2\gamma^4}\Bigl(\log^4m+\Bigl(\frac{1-p}{p}\Bigr)^4\log^4n\\&\textstyle +\log\frac1\delta\log^2m+\log\frac1\delta\\&\textstyle \qquad\cdot\Bigl(\frac{1-p}{p}\Bigr)^4\log^2n\Bigr)\Bigr)\end{aligned}$\\
\midrule
$L_p$ Norm Approxi\-mation, $p\in[1,2]$ & \cite{blasiok_et_al:LIPIcs.APPROX-RANDOM.2017.32} & N/A & $0$ & $O\Bigl(\frac{\log m}{\gamma^2}\Bigl(\log\log m+\log\frac1\gamma+\log\frac1\delta\Bigr)\Bigr)$\\*[3pt]
 & This Work & $(\varepsilon,\delta)$ & $\widetilde O\Bigl(\frac{1}{\varepsilon\gamma}\log m\cdot\log\frac{m}{\delta}\Bigr)$ & $\begin{aligned}[t]&\textstyle \widetilde O\Bigl(\frac{1}{\varepsilon^2\gamma^4}\Bigl(\log^5m\log\frac1\delta\\&\textstyle +\log^3m\log^3\frac1\delta\Bigr)\Bigr)\end{aligned}$\\
\midrule
$L_p$ Norm Approxi\-mation for $p>2$ & \cite{10.1145/1060590.1060621} & N/A & $0$ & $\widetilde O\Bigl(\frac1{\gamma^3}m^{1-\frac2p}\log\frac1\delta\cdot\operatorname{poly}\Bigl(\frac1\gamma,\log\frac1\delta\Bigr)\Bigr)$\\*[3pt]
 & This Work & $(\varepsilon,\delta)$ & $\widetilde O\Bigl(\frac{1}{\varepsilon\gamma}\log m\cdot\log\frac{m}{\delta}\Bigr)$ & $\widetilde O\Bigl(n^{1-\frac2p}\cdot\operatorname{poly}\Bigl(\frac1\varepsilon,\frac1\gamma,\log m,\log\frac1\delta\Bigr)\Bigr)$\\
\midrule
Longest Increasing Subsequence & \cite{GopalanJayramKrauthgamerKumar2007} & N/A & $0$ & $\widetilde O\bigl((m/\gamma)^{\frac12}\bigr)$\\*[3pt]
 & This Work & $(\varepsilon,\delta)$ & $\widetilde O\Bigl(\frac{1}{\varepsilon\gamma}\log m\cdot\log\frac{m}{\delta}\Bigr)$ & $\widetilde O\Bigl(\frac1\gamma\cdot\Bigl(\frac{m\log\frac1\delta}{\varepsilon}\Bigr)^{\frac12}\Bigr)$\\
\midrule
Linear Regression & \cite{braverman2020near} & N/A & $0$ & $O\Bigl(\frac{d^2}{\gamma^2}\cdot\log m\log\kappa_A\Bigr)$\\*[3pt]
 & This Work & $(\varepsilon,\delta)$ & $\widetilde O\Bigl(\frac m{\varepsilon\gamma}\log m\cdot\log\frac m\delta\Bigr)$ & $\begin{aligned}[t]&\textstyle \widetilde O\Bigl(\frac{d^2}{\varepsilon^2\gamma^4}\log\kappa_A\cdot\Bigl(\log^4m\\&\textstyle +\log^2m\log^2\frac1\delta\Bigr)\Bigr)\end{aligned}$\\
\midrule
Low-Rank Approxi\-mation & \cite{braverman2020near} & N/A & $0$ & $O\Bigl(\frac{kd}{\gamma^2}\cdot\log m\log^2\kappa_A\Bigr)$\\*[3pt]
 & This Work & $(\varepsilon,\delta)$ & $\widetilde O\Bigl(\frac{1}{\varepsilon\gamma}\log m\cdot\log\frac{m}{\delta}\Bigr)$ & $\begin{aligned}[t]&\textstyle \widetilde O\Bigl(\frac{kd}{\varepsilon^2\gamma^4}\log^2\kappa_A\cdot\Bigl(\log^5m\\&\textstyle +\log^3m\log^2\frac1\delta\Bigr)\Bigr)\end{aligned}$\\
\midrule
$(k,z)$-Mean Clustering & \cite{CohenAddadWoodruffZhou2023} & N/A & $0$ & $\widetilde O\Bigl(\frac{k}{\gamma^2}\Bigr)\cdot\min\Bigl\{\frac1{\gamma^2},k\Bigr\}$\\*[3pt]
 & This Work & $(\varepsilon,\delta)$ & $\begin{aligned}[t]&\textstyle \widetilde O\Bigl(\frac1{\varepsilon\gamma}\log m\cdot\log\frac m\delta\Bigr)\\&\textstyle \cdot\operatorname{poly}(m)\end{aligned}$ & $\widetilde O\Bigl(\frac{k\log^2m\log^2\frac m\delta}{\varepsilon^2\gamma^4}\Bigr)\cdot\min\Bigl\{\frac1{\varepsilon^z\gamma^{2z}},k\Bigr\}$\\
\midrule
Union Volume Estimation & \cite{bhore2026dynamic} & N/A & $0$ & $O\Bigl(\frac1{\gamma^2}\log^2m\Bigr)$\\*[3pt]
 & This Work & $(\varepsilon,\delta)$ & $\widetilde O\Bigl(\frac{V_{\max}}{\varepsilon\gamma}\cdot\log m\log\frac m\delta\Bigr)$ & $\widetilde O\Bigl(\frac{\log^4m\log^2\frac m\delta}{\varepsilon^2\gamma^4}\Bigr)$\\
\end{longtable}
\endgroup
In \Cref{tab:private-comparison}, we compare our framework with application-specific private continual-release algorithms. Our pure $\varepsilon$-private guarantees are justified in \Cref{sec:pure-dp}. For distinct elements, $L_p$ norms, and other statistics, our framework provides comparable polylogarithmic error and space guarantees. The benefit of our framework is its black-box nature. Rather than designing an individual private algorithm for each statistic, we transform a non-private continual approximation algorithm whenever the conditions of \Cref{thm:main} are met. Note that the algorithm of \cite{aamand2026skirting} also supports turnstile streams, which permit both insertions and deletions.
\begingroup
\small
\renewcommand{\arraystretch}{1.25}
\setlength{\tabcolsep}{3pt}
\begin{longtable}{@{}>{\raggedright\arraybackslash}p{\dimexpr 0.13\linewidth-2\tabcolsep\relax}>{\raggedright\arraybackslash}p{\dimexpr 0.1\linewidth-2\tabcolsep\relax}>{\raggedright\arraybackslash}p{\dimexpr 0.105\linewidth-2\tabcolsep\relax}>{\raggedright\arraybackslash}p{\dimexpr 0.27\linewidth-2\tabcolsep\relax}>{\raggedright\arraybackslash}p{\dimexpr 0.395\linewidth-2\tabcolsep\relax}@{}}
\caption{Comparison with application-specific algorithms on insertion-only streams. All rows have multiplicative factor $1+O(\gamma)$, except the longest-increasing-subsequence result of \cite{bonomi2016lis}, which has factor $O(\log m)$. For pure-private rows, $\delta$ is an accuracy-failure parameter.}\label{tab:private-comparison}\\
\toprule
Application & Reference & Privacy & Additive error & Space complexity\\
\midrule
\endfirsthead
\caption[]{Comparison with application-specific algorithms on insertion-only streams (continued).}\\
\toprule
Application & Reference & Privacy & Additive error & Space complexity\\
\midrule
\endhead
\midrule
\multicolumn{5}{r}{Continued on next page}\\
\endfoot
\bottomrule
\endlastfoot
Distinct Elements & \cite{epasto2023differentially} & $\varepsilon$ & $O\Bigl(\frac{\log^2(m/\delta)}{\varepsilon\gamma^2}\Bigr)$ & $\operatorname{poly}\Bigl(\frac{\log(m/\delta)}{\gamma\min\{\varepsilon,1\}}\Bigr)$\\*[3pt]
 & This Work & $(\varepsilon,\delta)$ & $\widetilde O\Bigl(\frac{1}{\varepsilon\gamma}\log m\cdot\log\frac{m}{\delta}\Bigr)$ & $\begin{aligned}[t]&\textstyle \widetilde O\Bigl(\frac{1}{\varepsilon^2\gamma^4}\Bigl(\log^4m\log\frac1\delta\\&\textstyle +\log^2m\log^3\frac1\delta\Bigr)\Bigr)\end{aligned}$\\*[3pt]
 & This Work & $\varepsilon$ & $\begin{aligned}[t]&\textstyle \widetilde O\Bigl(\frac{\log^3m}{\varepsilon\gamma^2}+\frac{\log m}{\varepsilon\gamma}\\&\textstyle \qquad\cdot\log\frac1{\varepsilon\delta}\Bigr)\end{aligned}$ & $\begin{aligned}[t]&\textstyle \widetilde O\Bigl(\frac{\log^8m}{\varepsilon^2\gamma^7}+\frac{\log^2m}{\varepsilon^2\gamma^4}\\&\textstyle \qquad\cdot\log^3\frac1{\varepsilon\delta}\Bigr)\end{aligned}$\\
\midrule
$L_p$ Norm Approxi\-mation, $p\in[1,2]$ & \cite{epasto2023differentially} & $\varepsilon$ & $\Bigl(\frac{\log(mn/\delta)}{\varepsilon\gamma}\Bigr)^{O(\max\{1,p\}/p)}$ & $\Bigl(\frac{\log(mn/\delta)}{\varepsilon\gamma}\Bigr)^{O(\max\{1,p\})}$\\*[3pt]
 & This Work & $(\varepsilon,\delta)$ & $\widetilde O\Bigl(\frac{1}{\varepsilon\gamma}\log m\cdot\log\frac{m}{\delta}\Bigr)$ & $\begin{aligned}[t]&\textstyle \widetilde O\Bigl(\frac{1}{\varepsilon^2\gamma^4}\Bigl(\log^5m\log\frac1\delta\\&\textstyle +\log^3m\log^3\frac1\delta\Bigr)\Bigr)\end{aligned}$\\*[3pt]
 & This Work & $\varepsilon$ & $\begin{aligned}[t]&\textstyle \widetilde O\Bigl(\frac{\log^3m}{\varepsilon\gamma^2}+\frac{\log m}{\varepsilon\gamma}\\&\textstyle \qquad\cdot\log\frac1{\varepsilon\delta}\Bigr)\end{aligned}$ & $\begin{aligned}[t]&\textstyle \widetilde O\Bigl(\frac{\log^9m}{\varepsilon^2\gamma^7}+\frac{\log^3m}{\varepsilon^2\gamma^4}\\&\textstyle \qquad\cdot\log^3\frac1{\varepsilon\delta}\Bigr)\end{aligned}$\\
\midrule
$L_p$ Norm Approxi\-mation for $p>2$ & \cite{epasto2023differentially} & $\varepsilon$ & $\Bigl(\frac{\log(mn/\delta)}{\varepsilon\gamma}\Bigr)^{O(\max\{1,p\}/p)}$ & $\begin{aligned}[t]&\textstyle n^{1-\frac2p}\\&\textstyle \cdot\Bigl(\frac{\log(mn/\delta)}{\varepsilon\gamma}\Bigr)^{O(\max\{1,p\})}\end{aligned}$\\*[3pt]
 & This Work & $(\varepsilon,\delta)$ & $\widetilde O\Bigl(\frac{1}{\varepsilon\gamma}\log m\cdot\log\frac{m}{\delta}\Bigr)$ & $\begin{aligned}[t]&\textstyle \widetilde O\Bigl(n^{1-\frac2p}\cdot\operatorname{poly}\Bigl(\frac1\varepsilon,\frac1\gamma,\log m,\\&\textstyle \log\frac1\delta\Bigr)\Bigr)\end{aligned}$\\*[3pt]
 & This Work & $\varepsilon$ & $\begin{aligned}[t]&\textstyle \widetilde O\Bigl(\frac{\log^3m}{\varepsilon\gamma^2}+\frac{\log m}{\varepsilon\gamma}\\&\textstyle \qquad\cdot\log\frac1{\varepsilon\delta}\Bigr)\end{aligned}$ & $\begin{aligned}[t]&\textstyle \widetilde O\Bigl(n^{1-\frac2p}\cdot\operatorname{poly}\Bigl(\frac1\varepsilon,\frac1\gamma,\log m,\\&\textstyle \log\frac1\delta\Bigr)\Bigr)\end{aligned}$\\
\midrule
$L_2$ Norm Approxi\-mation & \cite{aamand2026skirting} & $(\varepsilon,\delta)$ & $O\Bigl(\frac{\log^2m\sqrt{\log(1/\delta)}}{\varepsilon\gamma^{3/2}}\Bigr)$ & $O\Bigl(\frac{\log^2m}{\gamma^2}\Bigr)$\\*[3pt]
 & This Work & $(\varepsilon,\delta)$ & $\widetilde O\Bigl(\frac{1}{\varepsilon\gamma}\log m\cdot\log\frac{m}{\delta}\Bigr)$ & $\begin{aligned}[t]&\textstyle \widetilde O\Bigl(\frac{1}{\varepsilon^2\gamma^4}\Bigl(\log^5m\log\frac1\delta\\&\textstyle +\log^3m\log^3\frac1\delta\Bigr)\Bigr)\end{aligned}$\\*[3pt]
 & This Work & $\varepsilon$ & $\begin{aligned}[t]&\textstyle \widetilde O\Bigl(\frac{\log^3m}{\varepsilon\gamma^2}+\frac{\log m}{\varepsilon\gamma}\\&\textstyle \qquad\cdot\log\frac1{\varepsilon\delta}\Bigr)\end{aligned}$ & $\begin{aligned}[t]&\textstyle \widetilde O\Bigl(\frac{\log^9m}{\varepsilon^2\gamma^7}+\frac{\log^3m}{\varepsilon^2\gamma^4}\\&\textstyle \qquad\cdot\log^3\frac1{\varepsilon\delta}\Bigr)\end{aligned}$\\
\midrule
Longest Increasing Subsequence & \cite{bonomi2016lis} & $\varepsilon$ & $O\Bigl(\frac{\log^{3/2}m}{\varepsilon}\cdot\log\frac m\delta\Bigr)$ & $O(m)$\\*[3pt]
 & This Work & $(\varepsilon,\delta)$ & $\widetilde O\Bigl(\frac{1}{\varepsilon\gamma}\log m\cdot\log\frac{m}{\delta}\Bigr)$ & $\widetilde O\Bigl(\frac1\gamma\cdot\Bigl(\frac{m\log\frac1\delta}{\varepsilon}\Bigr)^{\frac12}\Bigr)$\\*[3pt]
 & This Work & $\varepsilon$ & $\begin{aligned}[t]&\textstyle \widetilde O\Bigl(\frac{\log^3m}{\varepsilon\gamma^2}+\frac{\log m}{\varepsilon\gamma}\\&\textstyle \qquad\cdot\log\frac1{\varepsilon\delta}\Bigr)\end{aligned}$ & $\begin{aligned}[t]&\textstyle \widetilde O\Bigl(\frac{\sqrt m\log m}{\sqrt\varepsilon\gamma^{3/2}}\\&\textstyle +\frac{\sqrt{m\log(1/(\varepsilon\delta))}}{\sqrt\varepsilon\gamma}\Bigr)\end{aligned}$\\
\end{longtable}
\endgroup

\subsection{Distinct Elements}
We now apply the framework on estimating the number of distinct elements in an insertion only stream. Let the stream be

$$D = (e_1, e_2, \cdots, e_m) \in [n]^m$$
where $n$ is the size of the universe. Define the distinct-elements function at time $i$ by

$$F_{0,i}(D) := |\{e_1, e_2, \cdots, e_i\}|.$$
We first prove the algorithms for Distinct Elements fit the requirements for our framework. 
\begin{lemma}\label{lem:distinctelem}
The function $F_0$ is non-negative and monotone on insertion-only streams, has sensitivity $\Delta_{F_{0}} = 1$, and has range bounded by $M = m$.
\end{lemma}
\begin{proof}
Adding a stream element can only increase the number of distinct elements so $F_{0}(D)$ is non-negative and monotone with respect to $i$. Additionally, one update can add at most one new distinct value, therefore, for neighboring streams $D\sim D'$,

$$\Delta_{F_0} = \left|F_{0,i}(D)-F_{0,i}(D')\right|\leq 1$$
for every time $i$. Finally, no prefix contains more than $m$ distinct elements from the universe, therefore $F_{0,i}(D) \leq m$ for all time $i$.
\end{proof}

Bar-Yossef et al.~\cite{646978.711822} provide a one-pass algorithm for distinct elements achieving a $(1 + \gamma)$-approximation of the number of distinct elements in a stream with probability $\delta$. 
Algorithms can be made accurate in the continual release setting by applying a union bound over the failure events at all possible release times. 
This effectively replaces factors of $\log\frac{1}{\delta}$ with $\log\frac{m}{\delta}$ in the space complexity. 
With this modification, Algorithm 1 requires $O(\frac{1}{\gamma^2}\log m \log\frac{m}{\delta})$ bits, and Algorithms 2 and 3 require $O((\frac{1}{\gamma^2}+\log m) \log\frac{m}{\delta})$ bits. These bounds were improved by Blasiok~\cite{10.1145/3309193} which provided a bound of $O(\log n+\frac{1}{\gamma^2}(\log\frac{1}{\delta}+\log\log n))$ bits, removing reliance on $m$ altogether.
\begin{theorem}
Let $\gamma\in(0,1]$ be the desired multiplicative accuracy, $\delta\in(0,1)$ be failure probability, and $\varepsilon>0$ be the privacy parameter. For the distinct-elements algorithm in \cite{10.1145/3309193}, there exists a private continual release algorithm for $F_0$ which releases $\bar{Y}_i$ at each time $i$ with the following guarantees

\begin{enumerate}
    \item \textbf{Privacy.} The complete released stream is $(\varepsilon,\delta)$-differentially private.

    \item \textbf{Accuracy.} With probability at least $1-\delta$, simultaneously for every time $i$,
    $$
    (1-\gamma)F_{0,i}(D)-\rho
    \leq
    \bar{Y}_i
    \leq
    (1+\gamma)F_{0,i}(D)+\rho,
    $$
    where
    $$
    \rho
    =
    \frac{36(\log{m} + 1)}{\varepsilon \gamma}
    \log\left(
    \frac{6(m+1)(\log{m} + 1)}{\gamma \delta}
    \right) +2.
    $$
    Equivalently, \textsc{PrivateStream} is $(\gamma,\rho,\delta)$-continually accurate.

    \item \textbf{Space.} Space complexity is at most $$O\left(\frac{\log^2 m \log^2\left( \frac{m}{\gamma\delta}\right)}{\varepsilon^2\gamma^4}\left(\log\frac{1}{\delta}+\log\log n\right)\right)$$ bits and equivalently, $\widetilde{O}\left(\frac{1}{\varepsilon^2\gamma^4}(\log^4m\log\frac{1}{\delta}+\log^2m\log^3\frac{1}{\delta})\right)$ bits.
\end{enumerate}
\end{theorem}
\begin{proof}
    By the structural properties established in \Cref{lem:distinctelem}, $F_{0}$ is non-negative, monotone, and insertion only, has sensitivity $\Delta_F = 1$, and range bounded by $M = m$. The privacy and accuracy guarantees therefore follow from \Cref{thm:main} by setting $\Delta_F = 1, M = m,$ and $\kappa = 0$.
    
    It remains to evaluate the space complexity which we measure in bits. The framework runs the base distinct elements algorithm with multiplicative approximation parameter
    
    $$\tau = \frac{\varepsilon\gamma^2}
    {288(\log m+1) \log\left( \frac{6(m+1)(\log m+1)}{\gamma\delta}\right)}$$
    and failure probability $\delta/2$.
    These parameters compound with the space requirement of the \cite{10.1145/3309193} algorithm, yielding,
    \begin{align*}
        O(\frac{1}{\tau^2}(\log\frac{2\log n}{\delta})+\log n)&=O\left(\frac{\left(288(\log m+1) \log\left( \frac{6(m+1)(\log m+1)}{\gamma\delta}\right)\right)^2}{\left(\varepsilon\gamma^2\right)^2}\left(\log\frac{2}{\delta}+\log\log n\right)+\log n\right)\\
        &=O\left(\frac{\log^2 m \log^2\left( \frac{m}{\gamma\delta}\right)}{\varepsilon^2\gamma^4}\left(\log\frac{1}{\delta}+\log\log n\right)+\log n\right).
    \end{align*}
    \Cref{thm:main} shows that \textsc{PrivateStream} additionally maintains $O(N)=O\left(\frac{\log m}{\gamma}\right)$ \textsc{AboveThreshold} instances, each storing a constant number of machine words: its threshold, clipping radius, noisy threshold, and halting flag. Every stored value has magnitude at most $\operatorname{poly}(nm)$ (thresholds are bounded by $M=m$, indices by $m$, and universe elements by $n$), so a machine word occupies $O(\log(nm))$ bits. The threshold instances therefore contribute $O(\frac{\log m}{\gamma}\log(nm))$ bits. Adding the base-algorithm and threshold costs gives
    $$O\left(\frac{\log^2 m \log^2\left( \frac{m}{\gamma\delta}\right)}{\varepsilon^2\gamma^4}\left(\log\frac{1}{\delta}+\log\log n\right)+\log n+\frac{\log m}{\gamma}\log(nm)\right)$$
    bits. Recalling that at most $m$ distinct elements ever appear, so $n=\operatorname{poly}(m)$ suffices, both the $\log n$ term and the threshold term $\frac{\log m}{\gamma}\log(nm)$ are dominated by the base-algorithm cost, and the total space is
    $$O\left(\frac{\log^2 m \log^2\left( \frac{m}{\gamma\delta}\right)}{\varepsilon^2\gamma^4}\left(\log\frac{1}{\delta}+\log\log n\right)\right)$$
    bits. Suppressing logarithmic factors yields $\widetilde{O}\left(\frac{1}{\varepsilon^2\gamma^4}(\log^4m\log\frac{1}{\delta}+\log^2m\log^3\frac{1}{\delta})\right)$ bits.
\end{proof}

\subsection{\texorpdfstring{$L_p$}{Lp} Norm Approximation, \texorpdfstring{$p\in(0,2]$}{p in (0,2]}}
Next, we apply the framework to the problem of estimating the $L_p$ norm of the stream, for $0<p\leq2$. Let the stream be $$D=(e_1,e_2,\dots,e_m)\in[n]^m$$ where $n\in\mathbb{N}$. Define $x^{(i)}\in\mathbb{R}^n$ to be the frequency vector representing the elements seen in the stream at time $i$. We aim to approximate $L_{p,i}:=\Vert x^{(i)}\Vert_p$ privately. 
\begin{lemma}\label{lem:lpnorm}
    For $p\geq1$, the function $L_p$ is non-negative and monotone on insertion only streams, has sensitivity bounded by $\Delta_{L_p}=1$, and has range bounded by $M=m$.
\end{lemma}
\begin{proof}
    Let $p\in[1,2]$, making $\Vert\cdot\Vert_p$ a true norm. Consider neighboring streams $D\sim D'$ and their respective frequency vectors $x^{(i)},x'^{(i)}$ at an arbitrary time $i\in [m]$. The difference between their $L_p$ norms can be bounded by the reverse triangle inequality. Recall $x^{(i)}$ and $x'^{(i)}$ differ by one fixed entry $d\in[n]$, with $|x^{(i)}_d-x'^{(i)}_d|=1$ and for all other $j\neq d$, $x^{(i)}_j=x'^{(i)}_j$.
    \begin{align*}
        \left|\lVert x^{(i)}\rVert_p-\lVert x'^{(i)}\rVert_p\right|\leq& \lVert x^{(i)}-x'^{(i)}\rVert_p\\
        =&\left(\sum_{j\in[n]}| x^{(i)}_j-x'^{(i)}_j|^p \right)^\frac{1}{p}\\
        =&\left(1^p\right)^\frac{1}{p}=1
    \end{align*}
    Likewise, this implies that for all $D$ and all $i\in[m]$, $\left| \lVert x^{(i)}\rVert_p-\lVert x^{(i-1)}\rVert_p\right|\leq1$. As for all streams $x^{(0)}=\mathbf{0}$, $\lVert x^{(i)}\rVert_p\leq i\leq m.$ Similarly, let $d$ be seen in the stream at time $i$.
    \begin{align*}
        \lVert x^{(i)}\rVert_p^p&=\sum_{j\in[n]}(x^{(i)}_j)^p \\
        &=\left[\sum_{j\in[n]}(x^{(i-1)}_j)^p\right] +(x^{(i)}_d)^p-(x^{(i)}_d-1)^p\\
        &\geq\lVert x^{(i-1)}\rVert_p^p,\\
    \end{align*}
    With the last inequality holding because $x^p$ is monotonically increasing for the domain of natural numbers. This implies $\lVert x^{(i)}\rVert_p\geq \lVert x^{(i-1)}\rVert_p$.
\end{proof}

\begin{lemma}
    For $p\in(0,1)$, the function $L_p$ is non-negative and monotone on insertion only streams, has sensitivity bounded by $\Delta_{L_p}=\min\{mn^\frac{1-p}{p},\frac{1}{p}m^{1-p}n^\frac{(1-p)^2}{p}\}$, and has range bounded by $M=mn^\frac{1-p}{p}$, which is polynomial in $m$, as $n=O(m)$
\end{lemma}
\begin{proof}
    Let $p\in(0,1)$. Again consider $D,D'$ neighboring datastreams. Without loss of generality, let $x'^{(i)}_d=x^{(i)}_d+1$. First, we will prove $\Vert x\Vert_p$ is strictly increasing with respect to any single index of $x$. Because $x^p$ is strictly increasing,
    \begin{align*}
        \Vert x^{(i)}\Vert_p^p&=\sum_{j=1}^n\left(x^{(i)}_j\right)^p\\
        &\leq\sum_{j=1}^n\left[\left(x^{(i)}_j\right)^p\right]-\left(x^{(i)}_d\right)^p+\left(x'^{(i)}_d\right)^p\\
        &=\Vert x'^{(i)}\Vert_p^p
    \end{align*}
    Next, we bound the sensitivity. The trivial bound is $\Delta_{L_p}\leq\Vert x_m\Vert_p$. We can attain a sometimes-better bound using the mean value theorem. If $\Vert x^{(i)}\Vert_p=0$, we see clearly, $\left|\Vert x^{(i)}\Vert_p-\Vert x'^{(i)}\Vert_p\right|=1$. Thus, assume that $\Vert x_i\Vert_p\geq 1$, i.e. there is some non-zero frequency entry, not necessarily $d$, at time $i$. Then for some $y\in \left(\sum_{j=1}^n\left(x^{(i)}_j\right)^p,\sum_{j=1}^n\left(x'^{(i)}_j\right)^p\right)$
    \begin{align*}
        \left|\Vert x^{(i)}\Vert_p-\Vert x'^{(i)}\Vert_p\right|
        &=\Vert x'^{(i)}\Vert_p-\Vert x^{(i)}\Vert_p\\
        &=\frac{1}{p}y^{\frac{1}{p}-1}\left(\sum_{j=1}^n\left(x'^{(i)}_j\right)^p-\sum_{j=1}^n\left(x^{(i)}_j\right)^p\right)\\
        &\leq\frac{1}{p}\left(\Vert x'^{(i)}\Vert_p^p\right)^{\frac{1}{p}-1}\left(\left(x_d^{(i)}+1\right)^p-\left(x_d^{(i)}\right)^p\right)\\
        &\leq\frac{1}{p}\Vert x^{(m)}\Vert_p^{1-p}\\
    \end{align*}
    Here, we have used the fact that $\sum_{j=1}^nx_j^p$ is strictly increasing with respect to $x_d$, the only index that is changing to bound $y$ by $\Vert x'^{(i)}\Vert_p^p$. We have bounded $\left(x_d^{(i)}+1\right)^p-\left(x_d^{(i)}\right)^p$ by looking at two cases:
    \begin{itemize}
        \item If $x_d^{(i)}=0$, then $\left(x_d^{(i)}+1\right)^p-\left(x_d^{(i)}\right)^p=1$ plainly.
        \item Otherwise, $x_d^{(i)}\geq1$. By the mean value theorem and concavity of $x^p$, $\left(x_d^{(i)}+1\right)^p-\left(x_d^{(i)}\right)^p\leq p\left(x_d^{(i)}\right)^{p-1}<1.$
    \end{itemize}
    Last, we bound the maximum value of $L_p$ with respect to $m$ and $n$. We claim, for stream of length $m$, the largest $\Vert x^{(m)}\Vert_p$ value is bounded by the $L_p$ value when, for all $j$, $x^{(m)}_j=\frac{m}{n}$. Consider our arbitrary frequency vector at time $m$, $x^{(m)}$, and compare it to the uniform frequency vector $x^*$. Note, $\Vert x^{(m)}\Vert_1=\Vert x^*\Vert_1=m$. By concavity of $x^p$,
    $$\Vert x^{(m)}\Vert_p^p=\sum_{j=1}^n\left(x^{(m)}_j\right)^p\leq n\left(\frac{m}{n}\right)^p=\Vert x^*\Vert_p^p.$$
    It follows that $L_{p,i}(D)\leq n^\frac{1-p}{p}m$
\end{proof}
\cite{blasiok_et_al:LIPIcs.APPROX-RANDOM.2017.32} provides a strong tracking for approximating the $L_p$ norm which provides an $(\gamma,\delta)$-approximation with space usage bounded by $O(\frac{\log m}{\gamma^2}(\log\log m+\log\frac{1}{\gamma}+\log\frac{1}{\delta}))$ space using a $p$-stable sketch. We privatize this algorithm resulting in the following guarantee. 
\begin{theorem}
Let $\gamma\in(0,1]$ be the desired multiplicative accuracy, $\delta\in(0,1)$ be failure probability, and $\varepsilon>0$ be the privacy parameter. Using the algorithm in \cite{blasiok_et_al:LIPIcs.APPROX-RANDOM.2017.32}, there exists a continual release algorithm which releases $\bar{Y}_i$ at each time $i$ approximating $L_{p,i}(D)$ with the following guarantees

\begin{enumerate}
    \item \textbf{Privacy.} The complete released stream is $(\varepsilon,\delta)$-differentially private.

    \item \textbf{Accuracy.} With probability at least $1-\delta$, simultaneously for every time $i$,
    $$
    (1-\gamma)F_i(D)-\rho
    \leq
    \bar{Y}_i
    \leq
    (1+\gamma)F_i(D)+\rho,
    $$
    where for $p\in[1,2],$
    $$
    \rho
    =
    \frac{36(\log {m} + 1)}{\varepsilon \gamma}
    \log{\left(\frac{6(m+1)(\log m+1)}{\gamma\delta}\right)}
    +2,
    $$
    and for $p\in(0,1)$ with $\Delta_{L_p}=\min\{\frac{1}{p}m^{1-p}n^{\frac{(1-p)^2}{p}},mn^{\frac{1-p}{p}}\},$
    $$
    \rho
    =
    \frac{36\Delta_{L_p}(\log {m} +\frac{1-p}{p}\log{n}+ 1)}{\varepsilon \gamma}
    \log{\left(\frac{6(m+1)(\log {m} +\frac{1-p}{p}\log{n}+ 1)}{\gamma\delta}\right)}
    +2.
    $$
    Equivalently, \textsc{PrivateStream} is $(\gamma,\rho, \delta)$-continually accurate.

    \item \textbf{Space.} The mechanism uses at most
    $$\widetilde{O}\left(
        \frac{1}
        {\varepsilon^2\gamma^4}\left(\log^5m\,\log\frac{1}{\delta}\,+\log^3m\,\log^3\frac{1}{\delta}\right)
        \right)$$
    space for $p\in[1,2]$ and $$\widetilde{O}\left(\frac{\log m}{\varepsilon^2\gamma^4}\left(\log^4m+\left(\frac{1-p}{p}\right)^4\log^4n+\log\frac{1}{\delta}\log^2m+\log\frac{1}{\delta}\left(\frac{1-p}{p}\right)^4\log^2n\right)\right)$$ space for $p\in(0,1)$
\end{enumerate}
\end{theorem}
\begin{proof}
    By the structural properties established in \Cref{lem:lpnorm}, $F_{p}$ is non-negative, monotone, and insertion only, has sensitivity $\Delta_F$ dependent on $p$, and range bounded by $M$, also dependent on $p$. The privacy and accuracy guarantees therefore follow from \Cref{thm:main} by setting placeholder $\Delta_F, M$ and setting $\kappa = 0$.
    
    It remains to evaluate the space complexity. The framework runs the base $L_p$ algorithm with multiplicative approximation parameter
    
    $$\tau = \frac{\varepsilon\gamma^2}
    {288(\log M+1) \log\left( \frac{6(m+1)(\log M+1)}{\gamma\delta}\right)}$$
    and failure probability $\delta/2$.
    These are used as parameters in the space requirement of the \cite{blasiok_et_al:LIPIcs.APPROX-RANDOM.2017.32} algorithm, yielding,
    \begin{align*}
        O\left(\frac{\log m}{\tau^2}
            \log\left(\frac{2\log m}{\tau\delta}\right)\right)
        &=O\Biggl(
            \frac{\left(288(\log M+1)
            \log\left(\frac{6(m+1)(\log M+1)}{\gamma\delta}\right)\right)^2}
                 {(\varepsilon\gamma^2)^2}
        \\
        &\qquad\cdot
            \log m
            \left(
                \log\frac{\log m}{\delta}
                +
                \log\frac{
                    288(\log M+1)
                    \log\left(\frac{6(m+1)(\log M+1)}{\gamma\delta}\right)}
                    {\varepsilon\gamma^2}
            \right)
        \Biggr)
        \\
        &=O\left(
        \frac{\log^2 M\,
        \log^2\left(\frac{mM}{\gamma\delta}\right)\log m}
        {\varepsilon^2\gamma^4}
        \log\left(
        \frac{
        \log m\,\log M\,
        \log\left(\frac{mM}{\gamma\delta}\right)
        }{
        \varepsilon\gamma^2\delta
        }
        \right)
        \right).
    \end{align*}
    \Cref{thm:main} shows that \textsc{PrivateStream} adds $O(N) = O\left(\frac{\log M}{\gamma}\right)$ space for its threshold instances. This is absorbed by the asymptotically larger cost of running the $L_p$ algorithm. We plug in $M=m$ for $p\in[1,2]$, to attain $O\left(
        \frac{\log^3m\,
        \log^2\left(\frac{m}{\gamma\delta}\right)}
        {\varepsilon^2\gamma^4}
        \log\left(
        \frac{
        \log^2 m\,
        \log\left(\frac{m}{\gamma\delta}\right)
        }{
        \varepsilon\gamma^2\delta
        }
        \right)
        \right)=\widetilde{O}\left(
        \frac{1}
        {\varepsilon^2\gamma^4}\left(\log^5m\,\log\frac{1}{\delta}\,+\log^3m\,\log^3\frac{1}{\delta}\right)
        \right).$ We plug in $M=mn^\frac{1-p}{p}$ for $p\in(0,1)$ to attain $\widetilde{O}\left(\frac{\log m}{\varepsilon^2\gamma^4}\left(\log^4m+\left(\frac{1-p}{p}\right)^4\log^4n+\log\frac{1}{\delta}\log^2m+\log\frac{1}{\delta}\left(\frac{1-p}{p}\right)^4\log^2n\right)\right)$
\end{proof}

\subsection{\texorpdfstring{$L_p$}{Lp} Norm Approximation for \texorpdfstring{$p>2$}{p>2}}
At $p>2$ there is a known transition from $L_p$ estimation having logarithmic dependence on $m$ and $n$, to having polynomial dependence on $n$. We use the same definitions of an input to the stream $D$ as in the section above. \cite{10.1145/1060590.1060621} provides a $\widetilde{O}(n^{1-\frac{2}{p}}\log\frac{1}{\delta}\cdot\operatorname{poly}(\frac{1}{\gamma}))$ algorithm for computing a $(\gamma,\delta)$-approximation at every time.
\begin{theorem}
For the algorithm in \cite{10.1145/1060590.1060621}, there exists a continual release algorithm for $L_p$ with the following guarantees

\begin{enumerate}
    \item \textbf{Privacy.} The complete released stream is $(\varepsilon,\delta)$-differentially private.

    \item \textbf{Accuracy.} With probability at least $1-\delta$, simultaneously for every time $i$,
    $$
    (1-\gamma)L_{p,i}(D)-\rho
    \leq
    \bar{Y}_i
    \leq
    (1+\gamma)L_{p,i}(D)+\rho,
    $$
    where
    $$
    \rho
    =
    \frac{36(\log {m} + 1)}{\varepsilon \gamma}
    \log{\left(\frac{6(m+1)(\log m+1)}{\gamma\delta}\right)}
    +2.
    $$
    Equivalently, \textsc{PrivateStream} is $(\gamma,\rho, \delta)$-continually accurate.

    \item \textbf{Space.} The mechanism uses space at most

    $$\widetilde{O}\Biggl(
            n^{1-\frac{2}{p}}\cdot\operatorname{poly}\left(\frac{1}{\varepsilon}, \frac{1}{\gamma},\log m, \log \frac{1}{\delta}\right)\Biggr)$$
\end{enumerate}
\end{theorem}
\begin{proof}
    By the structural properties established in \Cref{lem:lpnorm}, $L_{p}$ is non-negative, monotone, and insertion only, has sensitivity $\Delta_{L_p} = 1$, and range bounded by $M = m$. The privacy and accuracy guarantees therefore follow from \Cref{thm:main} by setting $\Delta_F = 1, M = m,$ and $\kappa = 0$.
    
    It remains to evaluate the space complexity. The framework runs the base $L_p$ algorithm with multiplicative approximation parameter
    
    $$\tau = \frac{\varepsilon\gamma^2}
    {288(\log m+1) \log\left( \frac{6(m+1)(\log m+1)}{\gamma\delta}\right)}$$
    and failure probability $\delta/2$. 
    These are inputs in the space requirement of the \cite{10.1145/1060590.1060621} algorithm, yielding,
    \begin{align*}
        \widetilde{O}\left(n^{1-\frac{2}{p}}\log\frac{2}{\delta}\cdot \operatorname{poly}\left(\frac{1}{\tau}\right)\right)
        &=\widetilde{O}\Biggl(
            n^{1-\frac{2}{p}}\log\frac{2}{\delta}\cdot\operatorname{poly}\left(\frac{288(\log m+1)
            \log\left(\frac{6(m+1)(\log m+1)}{\gamma\delta}\right)}
                 {\varepsilon\gamma^2}\right)\Biggr)\\
        &=\widetilde{O}\Biggl(
            n^{1-\frac{2}{p}}\cdot\operatorname{poly}\left(\frac{1}{\varepsilon}, \frac{1}{\gamma},\log m, \log \frac{1}{\delta}\right)\Biggr).
    \end{align*}
    \Cref{thm:main} shows that \textsc{PrivateStream} adds $O(N) = O\left(\frac{\log m}{\gamma}\right)$ space for its threshold instances. This is absorbed by the asymptotically larger cost of running the $L_p$ algorithm.
\end{proof}
\subsection{Longest Increasing Subsequence}
For a stream $D = (e_1, \cdots, e_m)$ over an ordered alphabet let $F_{\mathrm{LIS},i}(D)$ be the length of the longest increasing subsequence of the prefix $D_{\leq i}.$
We first prove the algorithms for Longest Increasing Subsequence fit the requirements for our framework.
\begin{lemma}
The function $F_{\mathrm{LIS}}$ is non-negative and monotone on insertion-only streams, has sensitivity bounded by $\Delta_{F_{\mathrm{LIS}}} = 1$, and has range bounded by $M = m$. 
\end{lemma}
\begin{proof}
Appending an element cannot shorten the longest increasing subsequence, so $F_{\mathrm{LIS},i}$ is nondecreasing in $i$.  Deleting one stream element removes at most one element from any increasing subsequence, so the length changes by at most one.  Any increasing subsequence of a length-$m$ prefix has length at most $m$.
\end{proof}
\begin{lemma}
\label{lem:lis-prefix}
The streaming algorithm of \cite{GopalanJayramKrauthgamerKumar2007} is online; after processing any prefix $D_{\le i}$, its maintained state depends only on that prefix. Consequently, for every $i$, the algorithm's current estimate is exactly the estimate it would output if the stream terminated after processing $D_{\le i}$. Therefore, for every prefix,

$$(1-\tau)F_{\mathrm{LIS},i}(D) \leq \widehat{F}_{\mathrm{LIS},i}(D) \leq F_{\mathrm{LIS},i}(D),$$
using
$\widetilde O\!\left(\sqrt{\frac{m}{\tau}}\right)$
space.
\end{lemma}
\begin{proof}
The algorithm processes the stream sequentially, updating its maintained data structure after each insertion using only the current state and the newly arrived element. At any time $i$, the maintained state is therefore identical to the state that would be obtained by executing the algorithm on the prefix $D_{\le i}$ alone. Applying Theorem~2.3 of \cite{GopalanJayramKrauthgamerKumar2007} to the prefix $D_{\le i}$ yields the stated approximation guarantee. The space bound is unchanged.
\end{proof}
\begin{theorem}
For the longest-increasing subsequence algorithm in \cite{GopalanJayramKrauthgamerKumar2007}, there exists a continual release algorithm for $F_{\mathrm{LIS}}$ with the following guarantees

\begin{enumerate}
    \item \textbf{Privacy.} The complete released stream is $(\varepsilon,\delta)$-differentially private.

    \item \textbf{Accuracy.} With probability at least $1-\delta$, simultaneously for every time $i$,
    $$
    (1-\gamma)F_{\mathrm{LIS},i}(D)-\rho
    \leq
    \bar{Y}_i
    \leq
    (1+\gamma)F_{\mathrm{LIS},i}(D)+\rho,
    $$
    where
    $$
    \rho
    =
    \frac{36(\log {m} + 1)}{\varepsilon \gamma}
    \log{\left(\frac{6(m+1)(\log m+1)}{\gamma\delta}\right)}
    +2.
    $$
    Equivalently, \textsc{PrivateStream} is $(\gamma,\rho,\delta)$-continually accurate.

    \item \textbf{Space.} Space complexity is at most $$\widetilde O\left(\frac{1}{\gamma}\sqrt{ \frac{288m(\log m+1) \log{\left(\frac{6(m+1)(\log m+1)}{\gamma\delta}\right)}}{\varepsilon}} + \frac{\log m+1}{\gamma}\right)$$
    In particular, suppressing logarithmic factors, the space complexity is
    $$\widetilde O\left(\frac{\sqrt {m\log\frac{1}{\delta}}}{\gamma\sqrt{\varepsilon}}
\right).$$
\end{enumerate}
\end{theorem}
\begin{proof}
By the structural properties established above, $F_{\mathrm{LIS}}$ is non-negative, monotone, and insertion only, has sensitivity $\Delta_F = 1$, and range bounded by $M = m$. The privacy and accuracy guarantees therefore follow from \Cref{thm:main} by setting $\Delta_F = 1, M = m,$ and $\kappa = 0$.

It remains to evaluate the space complexity. The framework runs the base LIS algorithm with multiplicative approximation parameter

$$\tau = \frac{\varepsilon\gamma^2}
{288(\log m+1) \log\left( \frac{6(m+1)(\log m+1)}{\gamma\delta}\right)}$$

and failure probability $\delta/2$. The algorithm of \cite{GopalanJayramKrauthgamerKumar2007} uses
$\widetilde O\left(\sqrt{\frac{m}{\tau}}\right)$ space for multiplicative parameter $\tau$  where the logarithmic dependence
on the failure probability is absorbed into the
$\widetilde O(\cdot)$ notation. Substituting the displayed value of $\tau$ gives 
\begin{align*}
\widetilde O\left(\sqrt{\frac{m}{\tau}}\right)
&=\widetilde O\left(\sqrt{ \frac{288m(\log m+1) \log{\left(\frac{6(m+1)(\log m+1)}{\gamma\delta}\right)}}{\varepsilon\gamma^2}}\right)\\
&=\widetilde O\left(\frac{1}{\gamma}\sqrt{ \frac{288m(\log m+1) \log{\left(\frac{6(m+1)(\log m+1)}{\gamma\delta}\right)}}{\varepsilon}}\right)
\end{align*}
\Cref{thm:main} shows that \textsc{PrivateStream} adds $O(N)
= O\left(\frac{\log m+1}{\gamma}\right)$ space for its threshold instances. Adding the base-algorithm and threshold costs gives

$$\widetilde O\left(\frac{1}{\gamma}\sqrt{ \frac{288m(\log m+1) \log{\left(\frac{6(m+1)(\log m+1)}{\gamma\delta}\right)}}{\varepsilon}} + \frac{\log m+1}{\gamma}\right).$$

Equivalently, this is
$$\widetilde O\left(\frac{\sqrt {m\log\frac{1}{\delta}}}{\gamma\sqrt{\varepsilon}}
\right).$$
\end{proof}
\subsection{Linear Regression}
We turn to the problem of estimating loss in linear regression. Each observation in the stream at time $i$ consists of a feature vector $a_i\in\mathbb{R}^d$ paired with a label $b_i\in \mathbb{R}$. Let $A_i\in\mathbb{R}^{i\times d}$ be the data matrix and $\vec{b}_i\in\mathbb{R}^i$ be the label vector at time $i$. The goal at each timestep is to find a vector in $\mathbb{R}^d$, such that $L_i:=\sum_{j\in[i]}(a_j^Tx-b_j)^2=\Vert A_ix-\vec{b}_i\Vert_2^2$ is minimized. Let $x_i$ be the optimal such vector at time $i$. While the sequence of vectors $x_1,\dots,x_m$ cannot be privatized by our framework because entries of $x_i$ are not monotonically increasing with respect to $i$, the loss function $L$ can be privatized. Additionally, assume that at all times, $b_i\in[c,C]$ for some $c,C\in \mathbb{R}$ where $c\leq0\leq C$. Let $R:=C-c=O(1)$.
\begin{lemma}
    The function $L$ is monotonically increasing and thus attains its maximum value at time $m$. $L$ and $\Delta_L$ are both upper bounded by $R^2m$.
\end{lemma}
\begin{proof}
    Consider, for the sake of contradiction, a case where $L_i>L_{i+1}$. Then, $\sum_{j\in[i]}(a_j^Tx_{i+1}-b_j)^2\leq \sum_{j\in[i+1]}(a_j^Tx_{i+1}-b_j)^2=L_{i+1}<L_i$, contradicting optimality of $x_i$, as $x_{i+1}$ yields a smaller loss at time $i$. Thus, $L$ is increasing.
    Next, we will bound loss. Consider baseline solution $\tilde{x}=\vec{0}$. This leads to a loss value $\tilde{L}_i:=\sum_{j\in[i]}b_j^2\leq R^2i$. As we choose $x_i$ to minimize loss, $L_i\leq\tilde{L}_i\leq R^2m$. Because loss is non-negative, $\Delta_L\leq R^2m$.
    
    \textit{Note:} these bounds are tight. Consider a stream where $c=0,d=R$, $a_1,\dots,a_{m-1}=[a]$ for some non-zero $a\simeq 0$, and $b_1,\dots,b_{m-1}=R$. For $i\in[m-1]$, we have $L_i=0$ with the solution $x_i=[\frac{R}{a}]$. Now, let $a_m=1$ and $b_m=0$. $L_m=(m-1)(ax-R)^2+x^2$. This is minimized at $x_m=\frac{(m-1)aR}{(m-1)a^2+1}$. As $a\to 0$, $x_m\to 0$, resulting in a loss approaching $(m-1)R^2$.
\end{proof}
\cite{braverman2020near} provides a continuous release algorithm for spectral approximation of a matrix at each time when a new row is observed. Their algorithm uses a technique introduced in \cite{cohen2016online} by constructing an online coreset approximating true data. It guarantees, for each $i\in[m],$ a matrix $M_i$, such that $(1-\gamma)\Vert A_ix\Vert_2\leq \Vert M_ix\Vert_2\leq(1+\gamma)\Vert A_ix\Vert_2$ for all $x\in \mathbb{R}^d$. This algorithm requires storing $O(\frac{d}{\gamma^2}\log m\log \kappa_A)$ rows, where $\kappa_A$ is the condition number of the stream. This results in $O(\frac{d^2}{\gamma^2}\log m\log \kappa_A)$ space. We use this algorithm to release approximate loss at every timestep, 
\begin{lemma}
    There is an algorithm that can be used to produce approximate loss, $\tilde{L}_i$, at each $i$, such that $(1-\gamma)L_i\leq \tilde{L}_i\leq(1+\gamma)L_i$ for true optimal loss $L_i$. The guarantee requires storing $O(\frac{d}{\gamma^2}\log m\log \kappa_A)$.
\end{lemma}
\begin{proof}
    Let $i\in[m]$, $A_i$ and $b_i$ be as defined above, and
    $B_i=
    \begin{bmatrix}
    A_i\ \vert \ b_i
    \end{bmatrix}.
    $
    By the guarantee given by Theorem 5.4 of \cite{braverman2020near}, we can produce an approximation $M_i\in \mathbb{R}^{k\times (d+1)}$ for $k=O(\frac{d}{\gamma^2}\log m\log \kappa_A)$, such that $(1-\frac{\gamma}{3})\Vert B_i x\Vert_2\leq \Vert M_ix\Vert_2\leq(1+\frac{\gamma}{3})\Vert B_ix\Vert_2$, and solve for 
    
    $$\tilde{x}_i
    = \operatorname*{argmin}\limits_{\substack{x\in\mathbb{R}^{d}}}
    \Vert M_i\begin{bmatrix}
    x \\
    -1
    \end{bmatrix}\Vert_2^2.$$
    
    This produces an optimal least-squares solution for the sketched instance. Furthermore, for $\gamma\in(0,1)$, as we are guaranteed, $(1-\gamma)\Vert B_i x\Vert^2_2\leq \Vert M_ix\Vert^2_2\leq(1+\gamma)\Vert B_ix\Vert^2_2$. Let $\tilde{A}_i\in\mathbb{R}^{k\times d}$ be the first $d$ columns of $M_i$ and $\tilde{b}_i\in\mathbb{R}^k$ be the last column. The optimality of $\tilde{x}_i$ and the true data's solution $x_i$ in conjunction with the approximation's guarantee implies,
    \begin{align*}
        &\Vert \tilde{A}_i\tilde{x}_i-\tilde{b}_i \Vert_2^2\leq \Vert \tilde{A}_ix_i-\tilde{b}_i \Vert_2^2\leq (1+\gamma)\Vert A_ix_i-b_i \Vert_2^2,\\
        &\Vert \tilde{A}_i\tilde{x}_i-\tilde{b}_i \Vert_2^2\geq (1-\gamma)\Vert A_i \tilde{x}_i-b_i \Vert_2^2 \geq(1-\gamma)\Vert A_i x_i-b_i \Vert_2^2 .
    \end{align*}
    This implies a $\gamma$-approximation of true loss.
\end{proof}
Now that we have established boundedness, monotonicity, sensitivity of loss, and accuracy and space requirements of the approximation algorithm, we give a privacy guarantee.
\begin{theorem}
For online least squares regression loss, there exists a private continual release algorithm approximating $L$ with the following guarantees.
\begin{enumerate}
    \item \textbf{Privacy.} The complete released stream is $(\varepsilon,\delta)$-differentially private.

    \item \textbf{Accuracy.} With probability at least $1-\delta$, simultaneously for every time $i$,
    $$
    (1-\gamma)F_i(D)-\rho
    \leq
    \bar{Y}_i
    \leq
    (1+\gamma)F_i(D)+\rho,
    $$
    where
    $$
    \rho
    =
    \frac{36\,R^2m}{\varepsilon \gamma}
    (\log{R^2m} + 1)
    \log\!\left(
    \frac{6(m+1)(\log{R^2m} + 1)}{\gamma \delta}
    \right) +2.
    $$
    Equivalently, \textsc{PrivateStream} is $(\gamma,\rho,\delta)$-continually accurate.

    \item \textbf{Space.} The mechanism uses space at most
    $$\widetilde{O}\left(\frac{d^2}{\varepsilon^2\gamma^4}\log\kappa_A\left(\log^4m+\log^2m\log^2\frac{1}{\delta}\right)\right).
    $$
\end{enumerate}
\end{theorem}
\begin{proof}
    The accuracy and privacy guarantees follow from general guarantees of our framework by setting $$
    \tau
    =
    \frac{\varepsilon\gamma^2}
    {288(\log{R^2m} + 1)\log\!\left(\frac{6(m+1)(\log{R^2m}+1)}{\gamma\delta}\right)},
    $$ as the underlying error parameter of the linear regression approximation algorithm. This leads to a space requirement of 
    \begin{align*}
        O(\frac{d^2}{\tau^2}\log m\log\kappa_A)&=O\left(\frac{d^2\left(288(\log{R^2m} + 1)\log\!\left(\frac{6(m+1)(\log{R^2m}+1)}{\gamma\delta}\right)\right)^2}{(\varepsilon\gamma^2)^2}\log m\log \kappa_A\right)\\
        &=O\left(\frac{d^2\log^2{m}\log^2\!\left(\frac{m}{\gamma\delta}\right)}{\varepsilon^2\gamma^4}\log m\log \kappa_A\right)\\
        &=\widetilde{O}\left(\frac{d^2}{\varepsilon^2\gamma^4}\log\kappa_A\left(\log^4m+\log^2m\log^2\frac{1}{\delta}\right)\right),
    \end{align*}
    suppressing $\log$ terms and having $R=O(1)$. \Cref{thm:main} shows \textsc{PrivateStream} adds $O(\frac{\log (R^2m+1)}{\gamma})=O(\frac{\log m}{\gamma})$, which is asymptotically small for $\varepsilon=O(1)$.
\end{proof}
\subsection{Low-Rank Approximation}
We now apply the framework to tracking the rank-$k$ reconstruction loss of an insertion-only stream of rows. At time $i$, the stream has revealed rows $a_1, \cdots, a_i \in \mathbb{R}^d$ which make up a data matrix $A_i \in \mathbb{R}^{i \times d}$. We assume throughout that the rows are bounded, $\| a_j\|_2 \leq R$ for all $j$, with $R = O(1)$. We write $(A_i)_k$ for the best rank-$k$ approximation of $A_i$ with the Frobenius norm. The statistic we release is the optimal rank-$k$ reconstruction loss of the prefix,

$$F_{\mathrm{LRA},i}(D) := \| A_i - (A_i)_k \|_F^2 = \min_P \| A_i - A_i P \|_F^2$$
where the minimum is over rank-$k$ orthogonal projections $P \in \mathbb{R}^{d \times d}$ where $P = P^\top = P^2$ and $\operatorname{rank}(P) = k$. While the optimal rank-$k$ subspace itself cannot be privatized by our framework, since its defining directions are not monotone in $i$, the scalar reconstruction loss $F_{\mathrm{LRA}}$ can be, as we now demonstrate.
\begin{lemma}
\label{lem:lowrank}
For any fixed $k\geq 0$, the function $F_{\mathrm{LRA}}$ is non-negative and monotone
on insertion-only streams, has sensitivity bounded by $\Delta_{F_{\mathrm{LRA}}}=R^2$,
and has range bounded by $M=R^2m$, which is polynomial in $m$ since $R=O(1)$.
\end{lemma}
\begin{proof}
Non-negativity is immediate, as $F_{\mathrm{LRA},i}(D)$ is a sum of squared singular values.

We next prove monotonicity. Fix a stream $D$ and a time $i$, and let $P^*$ be an optimal rank-$k$ projection for $A_{i+1}$, so that $F_{\mathrm{LRA},i+1}(D)=\| A_{i+1}-A_{i+1}P^*\|_F^2$. Since $A_{i+1}=A_i\circ a_{i+1}$, appending the row $a_{i+1}$ only adds a non-negative term, 
\begin{align*}
F_{\mathrm{LRA},i+1}(D) &=\| A_i-A_iP^*\|_F^2+\| a_{i+1}(I-P^*)\|_2^2\\
&\geq\| A_i-A_iP^*\|_F^2\\
&\geq\min_P\| A_i-A_iP\|_F^2\\
&= F_{\mathrm{LRA},i}(D). 
\end{align*}
Thus $F_{\mathrm{LRA},i}$ is non-decreasing in $i$, and thus monotone.

We now bound the sensitivity. Let $D\sim D'$ be neighboring streams. Without loss of generality
$A_i'=A_i\circ a$ consists of the rows of $A_i$ together with one additional row $a$ with $\|a\|_2\leq R$. If the differing coordinate exceeds $i$, then $A_i=A_i'$ and the two losses are the same. Let $P^*$ be an optimal rank-$k$ projection for $A_i$. Evaluating the same projection on $A_i'$,
\begin{align*}
F_{\mathrm{LRA},i}(D') &\leq \| A_i'-A_i'P^* \|_F^2\\ 
&=\| A_i-A_iP^*\|_F^2+\| a(I-P^*)\|_2^2\\
&=F_{\mathrm{LRA},i}(D)+\| a(I-P^*)\|_2^2\\
&\leq F_{\mathrm{LRA},i}(D)+R^2,\end{align*}
using $\| a(I-P^*)\|_2\leq\| a\|_2\leq R$, since $I-P^*$ is an orthogonal projection. Combining this with monotonicity under insertion, $F_{\mathrm{LRA},i}(D')\geq F_{\mathrm{LRA},i}(D)$, this yields $0\leq F_{\mathrm{LRA},i}(D')-F_{\mathrm{LRA},i}(D)\leq R^2$, so $\Delta_{F_{\mathrm{LRA}}}=R^2$.

Finally, for any rank-$k$ orthogonal projection $P$,
$$\|A_i(I-P)\|_F^2\le\|A_i\|_F^2\le iR^2\le mR^2,$$
so the optimum obeys the same range bound.
\end{proof}
\begin{theorem}
For the online projection-cost preservation algorithm of \cite{braverman2020near},
there exists a private continual release algorithm for $F_{\mathrm{LRA}}$ with the
following guarantees.
\begin{enumerate}
    \item \textbf{Privacy.} The complete released stream is
    $(\varepsilon,\delta)$-differentially private.

    \item \textbf{Accuracy.} With probability at least $1-\delta$, simultaneously for
    every time $i$,
    $$
    (1-\gamma)F_i(D)-\rho
    \leq
    \bar{Y}_i
    \leq
    (1+\gamma)F_i(D)+\rho,
    $$
    where
    $$
    \rho
    =
    \frac{36\,R^2}{\varepsilon\gamma}
    (\log{(R^2m)}+1)
    \log\!\left(
    \frac{6(m+1)(\log{(R^2m)}+1)}{\gamma\delta}
    \right)+2.
    $$
    Equivalently, \textsc{PrivateStream} is $(\gamma,\rho,\delta)$-continually accurate.

    \item \textbf{Space.} The mechanism uses at most
    $$
    O\!\left(
        \frac{kd\,\log^3 m\,\log^2\!\left(\frac{m}{\gamma\delta}\right)\log^2\kappa_A}
             {\varepsilon^2\gamma^4}
    \right)
    $$
    words of memory. 
    Equivalently, this is $\widetilde{O}\left(\frac{kd}{\varepsilon^2\gamma^4}\log^2\kappa_A\left(\log^5m+\log^3m\log^2\frac{1}{\delta}\right)\right).$
\end{enumerate}
\end{theorem}
\begin{proof}
By \Cref{lem:lowrank}, $F_{\mathrm{LRA}}$ is non-negative, monotone, and insertion only, has sensitivity $\Delta_F=R^2$, and range bounded by $M=R^2m$. The base algorithm of \cite{braverman2020near} is a tunable
continual approximation algorithm with additive parameter $\kappa=0$. Thus, the privacy and accuracy guarantees follow from \Cref{thm:main} by setting $\Delta_F=R^2$,
$M=R^2m$, and $\kappa=0$; substituting these values into the general expression for $\rho$ (whose additive term is $2\kappa+2=2$) gives the stated additive error.

We now evaluate the space complexity. The framework runs the base projection-cost
preservation algorithm with multiplicative approximation parameter
$$\tau=\frac{\varepsilon\gamma^2}{288(\log(R^2m)+1)\log\!\left(\frac{6(m+1)(\log(R^2m)+1)}{\gamma\delta}\right)}$$
and failure probability $\delta/2$. Substituting this $\tau$ into the base space bound
$O\!\left(\frac{kd}{\tau^2}\log m\log^2\kappa_A\right)$ yields
\begin{align*}
O\!\left(\frac{kd}{\tau^2}\log m\log^2\kappa_A\right)
&=O\!\left(\frac{kd\left(288(\log(R^2m)+1)\log\!\left(\frac{6(m+1)(\log(R^2m)+1)}{\gamma\delta}\right)\right)^2}{(\varepsilon\gamma^2)^2}\log m\log^2\kappa_A\right)\\
&=O\!\left(\frac{kd\,\log^2 m\,\log^2\!\left(\frac{m}{\gamma\delta}\right)}{\varepsilon^2\gamma^4}\log m\log^2\kappa_A\right)\\
&=O\!\left(\frac{kd\,\log^3 m\,\log^2\!\left(\frac{m}{\gamma\delta}\right)\log^2\kappa_A}{\varepsilon^2\gamma^4}\right).
\end{align*}
\Cref{thm:main} shows that \textsc{PrivateStream} adds
$O(N)=O\!\left(\frac{\log(R^2m)+1}{\gamma}\right)=O\!\left(\frac{\log m}{\gamma}\right)$
space for its threshold instances, which is absorbed by the asymptotically larger cost
of running the base algorithm. This is equivalently $\widetilde{O}\left(\frac{kd}{\varepsilon^2\gamma^4}\log^2\kappa_A\left(\log^5m+\log^3m\log^2\frac{1}{\delta}\right)\right).$
\end{proof}

\subsection{\texorpdfstring{$(k,z)$}{(k,z)}-clustering}
We now apply the framework to the reconstruction loss of Euclidean ($k$,$z$)-clustering. Here each stream update is an observation $a_i \in [\Delta]^d$, and our goal is to track the the cost of the best set of $k$ centers on the points so far. For a set of points $Y$ and set of centers $C \subset \mathbb{R}^{d}$ with $|C| \leq k$, we write $\mathrm{cost}(Y,C) := \sum_{a \in Y} \min_{c \in C} \| a - c \|_2^z$, and for a stream $D = (a_1, \cdots, a_m)$ we let $X_i = \{a_1, \cdots, a_i\}$ denote the multiset of the first $i$ observations. The statistic we release is the optimal reconstruction loss of this prefix,

$$F_{(k,z),i}(D) := \min_{C \subset \mathbb{R}^d, |C| \leq k} \sum_{j=1}^{i} \min_{c \subset C} \| a_j - c \|_2^z = \min_{C \subset \mathbb{R}^d, |C| = k} \mathrm{cost}(X_i,C)$$

The two canonical special cases of $(k,z)$-clustering are $k$-medians ($z = 1$) and $k$-means ($z = 2$), but our framework applies for all $z \geq 1$. We make the assumption that the aspect ratio is bounded $(\Delta = \mathrm{poly}(m))$ with $d$ and $z$ constant, as in the streaming clustering literature. \cite{CohenAddadWoodruffZhou2023} provides a $\widetilde{O}(\frac{dk}{\gamma^2})(2^{z\log z})\min\{\frac{1}{\gamma^z},k\}\cdot\operatorname{poly}(\log\log n,\Delta)$ space requirement, which we will use as our base algorithm.
We first verify the reconstruction loss meets the structural requirements of our framework.

\begin{lemma} 
\label{lem:kzcluster}
For any constant $z \geq 1$, the function $F_{(k,z)}$ is non-negative and monotone on insertion-only streams, has sensitivity bounded by $\Delta_{F_{(k,z)}} = (\sqrt{d}\Delta)^z$, and has range bounded by $M = m(\sqrt{d}\Delta)^z$. This assumption keeps the range polynomial in $m$.

\end{lemma}
\begin{proof}
Each term $\| a_j - c\|_2^z \geq 0$, so every clustering cost is non-negative, and thus $F_{(k,z),i}(D)$ is as well.

We now move to demonstrating monotonicity. Fix a stream $D$ and let $C^*$ be an optimal set of centers for the prefix $X_i$, so that $F_{(k,z),i}(D) = \text{cost}(X_i, C^*)$. Since every element of the sum is non-negative, dropping the $i$-th observation only decreases the cost and the resulting cost is at least the optimum over $X_{i-1}$:
$$F_{(k,z),i}(D)=\sum_{j=1}^{i}\min_{c\in C^*}\| a_j-c\|_2^{z}\ \geq\ \sum_{j=1}^{i-1}\min_{c\in C^*}\| a_j-c\|_2^{z}\ \geq\ \min_{|C|=k}\mathrm{cost}(X_{i-1},C)=F_{(k,z),i-1}(D).$$
Hence, $F_{(k,z),i-1}$ is non-decreasing and monotone in $i$.

We now move to bounding the sensitivity. Choose two neighboring streams $D \sim D'$ and let $D'$ contain an additional point $a$. Choose an optimal clustering of the smaller prefix. For each nonempty cluster, its single-center objective is convex in the center for $z\ge1$, so a minimizer can be chosen in the convex hull of that cluster, and hence in $[\Delta]^d$. Thus we may take an optimal center set $C^* \subseteq[\Delta]^d$. Reusing $C^*$ after inserting $a$ gives
$$F_{(k,z),i}(D') \leq F_{(k,z),i}(D)+\min_{c\in C^*}\|a-c\|_2^z
\le F_{(k,z),i}(D)+(\sqrt d\,\Delta)^z.$$
Together with monotonicity, this proves the stated sensitivity bound.

We now move to bounding the range. Evaluating any fixed single-center-per-point upper bound, for every $C$ we have $\mathrm{cost}(X_i,C)\leq\sum_{j=1}^{i}(\sqrt{d}\,\Delta)^{z}\leq m\,(\sqrt{d}\,\Delta)^{z}$, and thus $F_{(k,z),i}(D)\leq m\,(\sqrt{d}\,\Delta)^{z}=M$. Under $\Delta=\mathrm{poly}(m)$ with $d,z$ constant, $M$ is polynomial in $m$.
\end{proof}
\begin{theorem}
For the $(k,z)$-clustering cost algorithm of \cite{CohenAddadWoodruffZhou2023}, there exists a private continual release algorithm for $F_{(k,z)}$ with the following guarantees.
\begin{enumerate}
    \item \textbf{Privacy.} The complete released stream is $(\varepsilon,\delta)$-differentially private.

    \item \textbf{Accuracy.} With probability at least $1-\delta$, simultaneously for every time $i$,
    $$
    (1-\gamma)F_i(D)-\rho
    \leq
    \bar{Y}_i
    \leq
    (1+\gamma)F_i(D)+\rho,
    $$
    where
    $$
    \rho
    =
    \frac{36\,(\sqrt{d}\,\Delta)^{z}}{\varepsilon \gamma}
    (\log{M} + 1)
    \log\!\left(
    \frac{6(m+1)(\log{M} + 1)}{\gamma \delta}
    \right) +2,
    \qquad M=m\,(\sqrt{d}\,\Delta)^{z}.
    $$
    Equivalently, \textsc{PrivateStream} is $(\gamma,\rho,\delta)$-continually accurate.

    \item \textbf{Space.} The mechanism uses at most
    $$
    \widetilde{O}\!\left(
        \frac{dk(2^{z\log{z}})\,(\log M+1)^{2}\log^{2}\!\left(\frac{m\log M}{\gamma\delta}\right)}{\varepsilon^{2}\gamma^{4}}
        \cdot
        \min\!\left\{
            \frac{(\log M+1)^{z}\log^{z}\!\left(\frac{m\log M}{\gamma\delta}\right)}{\varepsilon^{z}\gamma^{2z}},\ k
        \right\}
    \right)\cdot\mathrm{poly}(\log\log(m\Delta))
    $$
    words of memory. Under $\Delta=\mathrm{poly}(m)$ with $d,z$ constant, this is
    $$\widetilde{O}\!\left(\frac{k\log^2 m\log^2\frac{m}{\delta} }{\varepsilon^{2}\gamma^{4}}\cdot\min\!\left\{\frac{1}{\varepsilon^{z}\gamma^{2z}},\,k\right\}\right)$$
\end{enumerate}
\end{theorem}
\begin{proof}
By the structural properties established in \Cref{lem:kzcluster}, $F_{(k,z)}$ is non-negative, monotone, and insertion only, has sensitivity $\Delta_F=(\sqrt{d}\,\Delta)^{z}$, and range bounded by $M=m\,(\sqrt{d}\,\Delta)^{z}$. The base algorithm of \cite{CohenAddadWoodruffZhou2023} is a tunable continual approximation algorithm with additive parameter $\kappa=0$. The privacy and accuracy guarantees therefore follow from \Cref{thm:main} by setting $\Delta_F=(\sqrt{d}\,\Delta)^{z}$, $M=m\,(\sqrt{d}\,\Delta)^{z}$, and $\kappa=0$; substituting these values into the general expression for $\rho$ (whose additive term is $2\kappa+2=2$) gives the stated additive error.

It remains to evaluate the space complexity. The framework runs the base clustering algorithm with multiplicative approximation parameter
$$\tau=\frac{\varepsilon\gamma^{2}}{288(\log M+1)\log\!\left(\frac{6(m+1)(\log M+1)}{\gamma\delta}\right)}$$
and failure probability $\delta/2$. Substituting this $\tau$ into the base space bound $\widetilde{O}\!\left(\frac{k}{\tau^{2}}\right)\cdot\min\!\left(\frac{1}{\tau^{z}},k\right)\cdot\mathrm{poly}(\log\log(m\Delta))$ and using
$$\frac{1}{\tau^{2}}=O\!\left(\frac{(\log M+1)^{2}\log^{2}\!\left(\frac{m\log M}{\gamma\delta}\right)}{\varepsilon^{2}\gamma^{4}}\right),
\qquad
\frac{1}{\tau^{z}}=O\!\left(\frac{(\log M+1)^{z}\log^{z}\!\left(\frac{m\log M}{\gamma\delta}\right)}{\varepsilon^{z}\gamma^{2z}}\right),$$
yields the displayed space bound. \Cref{thm:main} shows that \textsc{PrivateStream} adds $O(N)=O\!\left(\frac{\log M+1}{\gamma}\right)$ space for its threshold instances, which is absorbed by the asymptotically larger cost of running the base clustering algorithm. Finally, under $\Delta=\mathrm{poly}(m)$ with $d,z$ constant, we have $\log M=O(\log m)$, and this becomes equivalent to
$$\widetilde{O}\!\left(\frac{k\log^2 m\log^2\frac{m}{\delta} }{\varepsilon^{2}\gamma^{4}}\cdot\min\!\left\{\frac{1}{\varepsilon^{z}\gamma^{2z}},\,k\right\}\right).$$
\end{proof}
\subsection{Union Volume Estimation}
We now apply the framework to union volume estimation. Here each stream update is a geometric object $X_i \subseteq [0, R]^d$. Writing $U_i := \bigcup_{j \leq i} X_j$ for the union of the first $i$ objects, the statistic we release is the volume, $F_{\mathrm{vol}, i}(D) := \operatorname{vol}(U_i).$
We assume that every admissible object $X$ satisfies
$V_{\min}\leq \operatorname{vol}(X)\leq V_{\max}$,
for public bounds $(V_{\min},V_{\max}>0)$, and write
$\Lambda:=\frac{V_{\max}}{V_{\min}}$ for the corresponding aspect ratio. We assume $R=\operatorname{poly}(m)$, $d$ is constant, and $\Lambda=\operatorname{poly}(m)$, consistent with the standard parameterization in the volume-estimation literature \cite{bhore2026dynamic}. We treat these global volume bounds, and hence $\Lambda$, as public parameters supplied independently of the private stream. Under these assumptions, the union volume is polynomially bounded in $m$. We first verify that it satisfies the remaining structural requirements of our framework.

\begin{lemma}
\label{lem:unionvol}
The function $F_{\mathrm{vol}}$ is non-negative and monotone on insertion-only streams, has sensitivity bounded by $\Delta_{F_{\mathrm{vol}}}=V_{\max}$, and has range bounded by $M=R^{d}$.
\end{lemma}
\begin{proof}
Non-negativity is immediate as volume is non-negative.

We now demonstrate monotonicity. Inserting an object $X_{i+1}$ only enlarges the union $U_i = \bigcup_{j\leq i}X_j\subseteq \bigcup_{j\leq i+1}X_j = U_{i+1}$, and volume is monotone under set inclusion, so $F_{\operatorname{vol},i}(D)=\operatorname{vol}(U_i)\leq\operatorname{vol}(U_{i+1})=F_{\mathrm{vol},i+1}(D)$. Hence $F_{\mathrm{vol},i}$ is non-decreasing and thus monotone in $i$.

We now bound the sensitivity. Let $D \sim D'$ be neighboring streams and W.L.G $D'$ contains one additional object $X$ at some time, so that the two unions at time $i$ differ by whether $X$ is included. By subadditivity and monotonicity of volume
$$0 \leq \operatorname{vol}(U_i \cup X) - \operatorname{vol}(U_i) \leq \operatorname{vol}(X) \leq V_{\max}$$
so $|F_{\mathrm{vol},i}(D)-F_{\mathrm{vol},i}(D')|\leq V_{\max}$, giving $\Delta_{F_{\mathrm{vol}}}=V_{\max}$.

Finally, the union is contained in the ambient domain, so $F_{\mathrm{vol},i}(D)=\operatorname{vol}(U_i) \leq \operatorname{vol}([0,R]^d)=R^{d}=M$. Under constant $d$ and $R=\operatorname{poly}(m)$, $M$ is polynomial in $m$.
\end{proof}
\begin{theorem}
For the streaming union volume estimation algorithm of \cite{bhore2026dynamic}, there exists a private continual release algorithm for $F_{\mathrm{vol}}$ with the following guarantees.
\begin{enumerate}
    \item \textbf{Privacy.} The complete released stream is $(\varepsilon,\delta)$-differentially private.

    \item \textbf{Accuracy.} With probability at least $1-\delta$, simultaneously for every time $i$,
    $$
    (1-\gamma)F_i(D)-\rho
    \leq
    \bar{Y}_i
    \leq
    (1+\gamma)F_i(D)+\rho,
    $$
    where
    $$
    \rho
    =
    \frac{36\,V_{\max}}{\varepsilon \gamma}
    (\log{M} + 1)
    \log\!\left(
    \frac{6(m+1)(\log{M} + 1)}{\gamma \delta}
    \right) +2,
    \qquad M=R^{d}.
    $$
    Equivalently, \textsc{PrivateStream} is $(\gamma,\rho,\delta)$-continually accurate. Under $R,\Lambda=\operatorname{poly}(m)$ with $d$ constant, the additive error is $\rho=\widetilde{O}(\frac{V_{\max}}{\varepsilon\gamma}\log m\log\frac{m}{\delta})$, which is small relative to the released volume once the union grows large compared to a single object.

    \item \textbf{Space.} The mechanism uses at most
    $$
    \widetilde{O}\!\left(
        \frac{\log^{2}m\,(\log M+1)^{2}\log^{2}\!\left(\frac{m\log M}{\gamma\delta}\right)}{\varepsilon^{2}\gamma^{4}}
    \right)
    $$
    words of memory. Under $R,\Lambda=\operatorname{poly}(m)$ with $d$ constant, this is
    $$
    \widetilde{O}\!\left(\frac{\log^4m\log^2\frac{m}{\delta}}{\varepsilon^{2}\gamma^{4}}\right).
    $$
\end{enumerate}
\end{theorem}
\begin{proof}
By the structural properties established in \Cref{lem:unionvol}, $F_{\mathrm{vol}}$ is non-negative, monotone, and insertion only, has sensitivity $\Delta_F=V_{\max}$, and range bounded by $M=R^{d}$. The streaming algorithm of \cite{bhore2026dynamic} (Theorem~2.2, specialized to insertion-only queries) maintains a $(1\pm\tau)$-approximation of the union volume in a single pass, and is therefore a tunable continual approximation algorithm for $F_{\mathrm{vol}}$ with additive parameter $\kappa=0$. As in the previous applications, running $O(\log\frac{m}{\delta})$ independent copies and taking the median, together with a union bound over the $m$ times, converts its per-query high-probability guarantee into simultaneous accuracy at every time. The privacy and accuracy guarantees therefore follow from \Cref{thm:main} by setting $\Delta_F=V_{\max}$, $M=R^{d}$, and $\kappa=0$; substituting these values into the general expression for $\rho$ (whose additive term is $2\kappa+2=2$) gives the stated additive error.

It remains to evaluate the space complexity. The framework runs the base algorithm with multiplicative approximation parameter
$$\tau=\frac{\varepsilon\gamma^{2}}{288(\log M+1)\log\!\left(\frac{6(m+1)(\log M+1)}{\gamma\delta}\right)}$$
and failure probability $\delta/2$. The algorithm of \cite{bhore2026dynamic} uses $O(\log\Lambda\,\log m/\tau^{2})$ words of memory, which is $O(\log^{2}m/\tau^{2})$ under $\Lambda=\operatorname{poly}(m)$; the median-of-copies amplification contributes a further $O(\log\frac{m}{\delta})$ factor absorbed into $\widetilde{O}(\cdot)$. Using
$$\frac{1}{\tau^{2}}=O\!\left(\frac{(\log M+1)^{2}\log^{2}\!\left(\frac{m\log M}{\gamma\delta}\right)}{\varepsilon^{2}\gamma^{4}}\right)$$
yields the displayed space bound. \Cref{thm:main} shows that \textsc{PrivateStream} adds $O(N)=O\!\left(\frac{\log M+1}{\gamma}\right)$ space for its threshold instances, which is absorbed by the base cost. Under $R,\Lambda=\operatorname{poly}(m)$ with $d$ constant we have $\log M=O(\log m)$, which yields $\widetilde{O}\!\left(\frac{\log^4m\log^2\frac{m}{\delta}}{\varepsilon^{2}\gamma^{4}}\right).$
\end{proof}

\section{Above Threshold}
\label{app:above-thresh}
\begin{algorithm}
\caption{\textsc{AboveThreshold}}
\label{alg:svt}
\begin{algorithmic}[1]
\Require Privacy Parameter $\varepsilon_{0}$, threshold $T$, sensitivity bound $\Delta$, query stream $q_1,q_2,\dots$
\Ensure Stream of responses $a_1, a_2, \dots$
\State $\widehat T \gets T+\operatorname{Lap}(2\Delta/\varepsilon_0)$
\For{each query $i$}
    \State $\nu_i \gets \operatorname{Lap}(4\Delta/\varepsilon_0)$
    \If{$q_i + \nu_i \geq \widehat{T}$}
        \State Output $a_i \gets \top$
        \State Halt
    \Else
        \State Output $a_i \gets \bot$
    \EndIf
\EndFor
\end{algorithmic}
\end{algorithm}
\begin{lemma}[Privacy of \textsc{AboveThreshold}]
\label{lem:privacy:svt-standard}
Let $q_1,\ldots,q_m$ be deterministic queries satisfying
$$|q_i(D)-q_i(D')|\leq \Delta$$
for every neighboring $D\sim D'$ and every $i$. If \textsc{AboveThreshold} uses threshold noise $\operatorname{Lap}(\frac{2\Delta}{\varepsilon_0})$ and query noise $\operatorname{Lap}(\frac{4\Delta}{\varepsilon_0})$, then \textsc{AboveThreshold} is $\varepsilon_0$-differentially private.
\end{lemma}
\begin{proof}
Fix neighboring databases $D\sim D'$. Let $A$ and $A'$ denote the
outputs of \textsc{AboveThreshold} on $D$ and $D'$, respectively. We
show that, for every possible output vector $a$,
$$\Pr[A=a]\leq \exp(\varepsilon_0)\Pr[A'=a].$$
First consider an output vector of the form
$$a=(\underbrace{\bot,\ldots,\bot}_{k-1},\top),$$
which indicates the noisy input is below the noisy threshold until time $k$. Let
$$\widehat{T}=T+\mu, \qquad \mu\sim\operatorname{Lap}\left(\frac{2\Delta}{\varepsilon_0}\right), \qquad \nu_i\sim\operatorname{Lap}\left(\frac{4\Delta}{\varepsilon_0}\right),$$
where $\nu_i$ denotes the noise added to query $q_i$. Fix arbitrary values of the query noises
$\nu_1,\ldots,\nu_{k-1}.$
We take probabilities only over the remaining randomness in
$\widehat{T}$ and $\nu_k$. Define
$g(D):=\max_{i<k}\{q_i(D)+\nu_i\}.$
Conditioned on the fixed values of
$\nu_1,\ldots,\nu_{k-1}$, the output is $a$ if and only if $\widehat{T}>g(D)$ and $q_k(D)+\nu_k\geq\widehat{T}.$
Because every query has sensitivity at most $\Delta$,
$|g(D)-g(D')|\leq\Delta.$
This implies,
\begin{align*}
g(D)
&=
\max_{i<k}\{q_i(D)+\nu_i\}\\
&\leq
\max_{i<k}\{q_i(D')+\nu_i+\Delta\}\\
&=
g(D')+\Delta,
\end{align*}
and the reverse inequality follows by symmetry.

Let $f_{\nu}$ denote the density of
$\operatorname{Lap}(\frac{4\Delta}{\varepsilon_0})$ and let $f_{\widehat{T}}$
denote the density of
$T+\operatorname{Lap}(\frac{2\Delta}{\varepsilon_0})$. Then
\begin{align*}
\Pr[A=a]
&=
\Pr\left[
\widehat{T}>g(D)
\text{ and }
q_k(D)+\nu_k\geq\widehat{T}
\right]\\
&=
\int_{-\infty}^{\infty}
\int_{-\infty}^{\infty}
f_{\nu}(v)f_{\widehat{T}}(t)
\mathbf{1}\left[
t\in\left(g(D),q_k(D)+v\right]
\right]
\,dt\,dv.
\end{align*}

To make a change of variables, let $t'=t-g(D)+g(D')$ and
$v'=v+q_k(D)-q_k(D')-g(D)+g(D').$ Observe, $t>g(D)\iff t'>g(D'),$ and
$q_k(D)+v\geq t\iff q_k(D')+v'\geq t'.$ Therefore, the indicator becomes
$\mathbf{1}\left[
t'\in\left(g(D'),q_k(D')+v'\right]
\right].$

To proceed with the change of variables, we bound $|t-t'|$ and $|v-v'|$ using the fact that $D,D'$ are neighboring inputs with bounded difference between their query responses.
$$
|t-t'|
=
|g(D)-g(D')|
\leq\Delta,
$$
and
\begin{align*}
|v-v'|
&=
\left|
q_k(D')-q_k(D)+g(D)-g(D')
\right|\\
&\leq
|q_k(D')-q_k(D)|
+
|g(D)-g(D')|\\
&\leq
2\Delta.
\end{align*}

For a Laplace random variable selected from distribution $\operatorname{Lap}(b)$, its density satisfies
$f(x+c)\leq \exp\left(\frac{|c|}{b}\right)f(x).$
Consequently, since the threshold noise has scale
$\frac{2\Delta}{\varepsilon_0}$ and query noise has scale
$4\Delta/\varepsilon_0$,
$$
f_{\widehat{T}}(t)
\leq
\exp\left(
\frac{\varepsilon_0|t-t'|}{2\Delta}
\right)
f_{\widehat{T}}(t')
\leq
\exp\left(\frac{\varepsilon_0}{2}\right)
f_{\widehat{T}}(t'),\qquad f_{\nu}(v)
\leq
\exp\left(
\frac{\varepsilon_0|v-v'|}{4\Delta}
\right)
f_{\nu}(v')
\leq
\exp\left(\frac{\varepsilon_0}{2}\right)
f_{\nu}(v').
$$
The change of variables has Jacobian one, so
\begin{align*}
\Pr[A=a]
&\leq
\exp(\varepsilon_0)
\int_{-\infty}^{\infty}
\int_{-\infty}^{\infty}
f_{\nu}(v')f_{\widehat{T}}(t')
\\
&\hspace{2cm}\cdot
\mathbf{1}\left[
t'\in\left(g(D'),q_k(D')+v'\right]
\right]
\,dt'\,dv'\\
&=
\exp(\varepsilon_0)\Pr[A'=a].
\end{align*}

Because this bound holds for every fixed choice of
$\nu_1,\ldots,\nu_{k-1}$, averaging over these noises preserves the
same inequality.

It remains to consider the vector $a=(\bot,\ldots,\bot),$
in which no query crosses the threshold. Fix all query noises
$\nu_1,\ldots,\nu_m$ and define
$g(D):=\max_{i\leq m}\{q_i(D)+\nu_i\}.$
The mechanism outputs this vector exactly when $\widehat{T}>g(D).$
As before, $|g(D)-g(D')|\leq\Delta.$ Now, only one source of randomness remains, so the change of variables $t'=t-g(D)+g(D')$ is sufficient.
The threshold-density bound gives
\begin{align*}
\Pr[A=a] &=\int_{-\infty}^{\infty} f_{\widehat{T}}(t) \mathbf{1}[t>g(D)]\,dt\\
&\leq \exp\left(\frac{\varepsilon_0}{2}\right) \int_{-\infty}^{\infty}
f_{\widehat{T}}(t') \mathbf{1}[t'>g(D')] \,dt'\\
&\leq \exp(\varepsilon_0)\Pr[A'=a].
\end{align*}
Averaging over the fixed query noises again preserves the inequality.

Thus, for every possible vector $a$, $\Pr[A=a] \leq \exp(\varepsilon_0)\Pr[A'=a].$
Summing this pointwise inequality over all vectors in an arbitrary
output event $S$ gives
$$\Pr[A\in S] \leq \exp(\varepsilon_0)\Pr[A'\in S].
$$
Therefore, \textsc{AboveThreshold} is
$\varepsilon_0$-differentially private.
\end{proof}
\begin{lemma}[Accuracy of AboveThreshold]\label{lem:AT-accuracy}
Let $q_1,\ldots,q_m$ be deterministic queries, and suppose that
AboveThreshold is run with threshold $T$, threshold noise
$$\widetilde T=T+\mu,
\qquad
\mu\sim\operatorname{Lap}\!\left(\frac{2\Delta}{\eps_0}\right),$$
and independent query noise
$$\nu_i\sim\operatorname{Lap}\!\left(\frac{4\Delta}{\eps_0}\right)
\qquad\text{for each }i\in[m].$$
For any $\delta^{\prime}\in(0,1)$, define
$$\alpha_{\mathrm{AT}}
:=\frac{6\Delta}{\eps_0}
\log\!\left(\frac{m+1}{\delta^{\prime}}\right).$$
Then, with probability at least $1-\delta^{\prime}$, both of the following
statements hold simultaneously:
\begin{enumerate}
    \item if AboveThreshold has not fired by time $i$, then
    $q_i(D)<T+\alpha_{\mathrm{AT}};$
    \item if AboveThreshold fires at time $k$, then
    $q_k(D)\geq T-\alpha_{\mathrm{AT}}.$
\end{enumerate}
\end{lemma}

\begin{proof}
The mechanism compares each noisy query $q_i(D)+\nu_i$ with the noisy
threshold $\widetilde T$.  Thus it fires at time $i$ precisely when
$q_i(D)+\nu_i\geq\widetilde T.$

Let
$h:=\log\!\left(\frac{m+1}{\delta^{\prime}}\right),$
so that $\alpha_{\mathrm{AT}}=6\Delta h/\eps_0$.  We first prove the two
accuracy statements under the event

$$\max_{i\in[m]}|\nu_i|+|\widetilde T-T|
\leq\alpha_{\mathrm{AT}}.$$

Suppose that the mechanism fires at time $k$.  By the definition of
AboveThreshold,
$q_k(D)+\nu_k\geq\widetilde T.$
It follows that
$$q_k(D)
\geq\widetilde T-\nu_k
\geq T-|\widetilde T-T|-|\nu_k|
\geq T-\alpha_{\mathrm{AT}}.$$
This proves the second statement.

Now suppose that the mechanism has not fired by time $i$.  In particular,
the query at time $i$ does not cross the noisy threshold, and hence
$q_i(D)+\nu_i<\widetilde T.$ Therefore,
$$q_i(D) <\widetilde T-\nu_i \leq T+|\widetilde T-T|+|\nu_i| \leq T+\alpha_{\mathrm{AT}}.$$
This proves the first statement.

It remains to bound the probability of the event used above.  Recall that if
$Y\sim\operatorname{Lap}(b)$, then
$\Pr[|Y|>tb]=e^{-t}.$
Since
$\widetilde T-T\sim\operatorname{Lap}\!\left(\frac{2\Delta}{\eps_0}\right),$
we have
$\Pr[
|\widetilde T-T|>\frac{2\Delta h}{\eps_0}
]=e^{-h}.$
Similarly, for every $i\in[m]$,
$\Pr\!\left[
|\nu_i|>\frac{4\Delta h}{\eps_0}
\right]=e^{-h}.$
A union bound over the threshold noise and the $m$ query-noise variables gives
$$\Pr\!\left[
|\widetilde T-T|>\frac{2\Delta h}{\eps_0}
\ \text{or}\
\max_{i\in[m]}|\nu_i|>\frac{4\Delta h}{\eps_0}
\right] \leq(m+1)e^{-h} =\delta^{\prime}.$$
Outside this failure event,
$$\max_{i\in[m]}|\nu_i|+|\widetilde T-T|
\leq\frac{4\Delta h}{\eps_0} +\frac{2\Delta h}{\eps_0}
=\frac{6\Delta h}{\eps_0}
=\alpha_{\mathrm{AT}}.$$
Therefore, both accuracy statements hold simultaneously with probability at
least $1-\delta^{\prime}$.
\end{proof}

\section{Acknowledgements}
ChatGPT (Sol 6.1) was used to polish grammar throughout the manuscript and in drafting Section~\ref{sec:applications-main} and the intro paragraphs of Section~\ref{app:applications}. The authors independently reviewed all revisions and take full responsibility for the manuscript.

\end{document}